\documentclass[12pt]{article}
\usepackage{amsmath,bm}
\usepackage{amssymb}
\usepackage{graphicx}
\usepackage{amsfonts}         
\usepackage{fancybox}   
\usepackage{yfonts} 
\usepackage{epsfig}
\usepackage{slashed}
\usepackage{subfigure}
\usepackage{wrapfig}
\usepackage{hyperref}
\usepackage{epstopdf}
\usepackage{color}
\usepackage{amscd}
\usepackage{tipa}

\usepackage{array}
\newcommand{\hl}[1]{{\color{red} #1}}

\renewcommand{\baselinestretch}{1.4}
\def\ads{\mbox{AdS}}

\def\comments#1{}

\def\p{\partial}

\def\Tr{{{\rm Tr }}}
\def\tr{{\rm tr }}

\def\Im{{\rm Im\hskip0.1em}}

\def\bra#1{{\langle}#1|}
\def\ket#1{|#1\rangle}

\def\vev#1{\langle{#1}\rangle}

\def\CF{{\cal F}}

\def\CN{{\cal N}}
\def\CO{{\cal O}}
\def\CR{{\cal R}}
\def\CH{{\cal H}}

\def\CL{{\cal L}}

\def\II{\relax{I\kern-.10em I}}

\def\IB{\relax{\rm I\kern-.18em B}}
\def\ID{\relax{\rm I\kern-.18em D}}
\def\IE{\relax{\rm I\kern-.18em E}}
\def\IF{\relax{\rm I\kern-.18em F}}
\def\IG{\relax\hbox{$\inbar\kern-.3em{\rm G}$}}
\def\IGa{\relax\hbox{${\rm I}\kern-.18em\Gamma$}}
\def\II{\relax{\rm I\kern-.18em I}}
\def\IK{\relax{\rm I\kern-.18em K}}

\def\inbar{\,\vrule height1.5ex width.4pt depth0pt}

\def\p{\partial}

\def\frac#1#2{{#1 \over #2}}

\newdimen\tableauside\tableauside=1.0ex
\newdimen\tableaurule\tableaurule=0.4pt
\newdimen\tableaustep
\def\phantomhrule#1{\hbox{\vbox to0pt{\hrule height\tableaurule width#1\vss}}}
\def\phantomvrule#1{\vbox{\hbox to0pt{\vrule width\tableaurule height#1\hss}}}
\def\sqr{\vbox{%
  \phantomhrule\tableaustep
  \hbox{\phantomvrule\tableaustep\kern\tableaustep\phantomvrule\tableaustep}%
  \hbox{\vbox{\phantomhrule\tableauside}\kern-\tableaurule}}}
\def\squares#1{\hbox{\count0=#1\noindent\loop\sqr
  \advance\count0 by-1 \ifnum\count0>0\repeat}}
\def\tableau#1{\vcenter{\offinterlineskip
  \tableaustep=\tableauside\advance\tableaustep by-\tableaurule
  \kern\normallineskip\hbox
    {\kern\normallineskip\vbox
      {\gettableau#1 0 }%
     \kern\normallineskip\kern\tableaurule}%
  \kern\normallineskip\kern\tableaurule}}
\def\gettableau#1 {\ifnum#1=0\let\next=\null\else
  \squares{#1}\let\next=\gettableau\fi\next}

 \def\eqnn#1{\xdef #1{(\secsym\the\meqno)}\writedef{#1\leftbracket#1}%
 \global\advance\meqno by1\wrlabeL#1}
 \def\eqna#1{\xdef #1##1{\hbox{$(\secsym\the\meqno##1)$}}
 \writedef{#1\numbersign1\leftbracket#1{\numbersign1}}%
 \global\advance\meqno by1\wrlabeL{#1$\{\}$}}
 \def\eqn#1#2{\xdef #1{(\secsym\the\meqno)}\writedef{#1\leftbracket#1}%
 \global\advance\meqno by1$$#2\eqno#1\eqlabeL#1$$}

\global\newcount\itemno \global\itemno=0

\def\itemaut#1{\global\advance\itemno by1\noindent\item{\the\itemno.}#1}

\def\del{\partial}

\def\({\left(}
\def\){\right)}

\def\eg{{\it e.g.}}
\def\ie{{\it i.e.}}

\newif{\ifeq}           
\eqtrue                 
\newcommand{\be}{\begin{equation}}
\newcommand{\ee}{\end{equation}}
\newcommand{\bea}{\begin{eqnarray}}
\newcommand{\eea}{\end{eqnarray}}
\newcommand{\bean}{\begin{eqnarray*}}
\newcommand{\eean}{\end{eqnarray*}}

\newcommand{\nn}{\nonumber}

\def\({\left(}
\def\){\right)}
\def\[{\left[}
\def\]{\right]}

\newcommand{\half}{\frac{1}{2}}

\renewcommand{\O}{{\cal O}}

\def\CO{\O}

\newcommand{\IR}{{\mathbb R}}

\def\ie{{\it i.e.}}

\newcommand{\lsim}{\,\raise.3ex\hbox{$<$\kern-.75em\lower1ex\hbox{$\sim$}}\,}
\newcommand{\gsim}{\,\raise.3ex\hbox{$>$\kern-.75em\lower1ex\hbox{$\sim$}}\,}

\def\p{\partial}

\newif{\ifeq}
\eqtrue

\begin{document}

\begin{titlepage}

\begin{flushright}
MIT-CTP/4067, NSF-KITP-09-134
\end{flushright}
\vfil

\begin{center}
{\huge Holographic duality}\\
\vskip.2in
{\huge with a view toward many-body physics}\\
\end{center}
\vfil
\begin{center}
{\large John McGreevy}\\
\vspace{1mm}
Center for Theoretical Physics, MIT,
Cambridge, Massachusetts 02139, USA\\
Kavli Institute for Theoretical Physics, Santa Barbara, California 93106-4030, USA 
{\tt mcgreevy at mit.edu}\\
\vspace{3mm}
\end{center}

\vfil

\begin{center}

{\large Abstract}
\end{center}

\noindent
These are notes based on a series of lectures given at the KITP
workshop {\it Quantum Criticality and the AdS/CFT Correspondence}
in July, 2009.
The goal of the lectures was to introduce condensed matter physicists 
to the AdS/CFT correspondence.
Discussion of string theory and of supersymmetry is avoided to the extent possible.

\vfill
\begin{flushleft}
September 2009, revised May 2010
\end{flushleft}
\vfil
\end{titlepage}
\newpage
\renewcommand{\baselinestretch}{1.1}  

\renewcommand{\arraystretch}{1.5}

\tableofcontents

\section{Introductory remarks}

My task in these lectures is to engender some understanding
of the following\\
{\bf Bold Assertion:} \\
(a) Some ordinary quantum field theories (QFTs) are secretly quantum theories of gravity.\\
(b) Sometimes the gravity theory is classical, and therefore 
we can use it to compute interesting observables of the QFT.

Part (a) is vague enough that it really just raises the questions:
`which QFTs?' and `what the heck is a quantum theory of gravity?'
Part (b) begs the question `when??!'

In trying to answer these questions, I have two conflicting goals:
On the one hand, I want to convince you that some 
statement along these lines is {\it true},
and on the other hand I want to convince you that it is {\it interesting}.
These goals conflict because our best evidence for the Assertion
comes with the aid of supersymmetry and complicated technology from string theory,
and applies to very peculiar theories which represent 
special cases of the correspondence, wildly over-represented in the literature
on the subject.
Since most of this technology is completely irrelevant for the applications
that we have in mind
(which I will also not discuss explicitly except
to say a few vague words at the very end), 
I will attempt to accomplish the first goal by way of showing
that the correspondence gives sensible answers to some interesting questions.
Along the way we will try to get a picture of its regime of validity.  

Material from other review articles,
including 
\cite{Horowitz:2006ct, Hartnoll:2009sz, 
Sachdev:2008ba,
Son:2007vk, Maldacena:2003nj, Aharony:1999ti, D'Hoker:2002aw},
has been liberally borrowed to construct these notes.
In addition, some of the tex source and most of the figures were pillaged from lecture notes from 
my class at MIT during Fall 2008 \cite{ocwref}, some of which were 
created by students in the class.
In particular I wish to thank
Christiana Athanasiou, 
Francesco D'Eramo,
Tom Faulkner, Tarun Grover,
Wing-Ko Ho, 
Vijay Kumar, Tongyan Lin, Daniel Park, and Brian Swingle; 
specific figure credits appear in the margins.
The discussion in section \ref{sec:nabil} follows
a talk given by Nabil Iqbal.
I'm grateful to Sean Hartnoll, Joe Polchinski, and Subir Sachdev
for helpful discussions and 
for giving me the opportunity to inflict 
my perspective on 
their workshop participants,
those participants for their many lively questions,
and to Pavel Kovtun for teaching me many of the ideas 
discussed herein.
Thanks also to Brian Swingle and T. Senthil for being
the first victims of many of these explanations, and to
Koushik Balasubramanian, Dan Freedman, Hong Liu, Kostas Skenderis and Erik Tonni for 
comments on the draft.
The title is adapted from 
\cite{Eisenbud}.
The selection of references was made based on perceived pedagogical value
and personal bias.

\section{Motivating the correspondence}

To understand what one might mean by a more precise version of the 
{\bf Bold Assertion} above, 
 we will follow 
 for a little while
 the interesting logic of \cite{Horowitz:2006ct},
which liberally uses hindsight, but 
does not use string theory.

Here are three facts which make the Assertion seem less unreasonable.

1) First we must define what we mean by a quantum gravity (QG).  As a working definition,
let's say that a QG is a quantum theory with a dynamical metric.
In enough dimensions, this usually means that there are local degrees of freedom.
In particular, linearizing equations of motion (EoM) for a metric usually reveals
a propagating mode of the metric, some spin-2 massless particle
which we can call a `graviton'.  

So at the least the assertion must mean that there is some spin-two graviton particle
that is somehow a composite object made of gauge theory degrees of freedom.
This statement seems to run afoul of the Weinberg-Witten no-go theorem,
which says:\\
{\bf Theorem} [Weinberg-Witten\cite{Jenkins:2006bz}]{\bf :}
A QFT with a Poincar\'e covariant
conserved stress tensor $T^{\mu\nu} $ forbids massless particles
of spin $j>1$ which carry momentum (\ie\ with $P^\mu = \int d^Dx T^{0\mu} \neq 0 $).

You may worry that the assumption of Poincar\'e invariance plays an important role
in the proof, but the set of QFTs to which the Bold Assertion applies includes
relativistic theories.

General relativity (GR) gets around this theorem because the total stress tensor (including the gravitational bit) vanishes by the metric EoM:
$ T^{\mu\nu} \propto {\delta S \over \delta g_{\mu\nu}} = 0 $.  (Alternatively, the `matter stress tensor,' which doesn't vanish,
is not general-coordinate invariant.)

Like any good no-go theorem, it is best considered a sign pointing away from wrong directions.
The loophole in this case is blindingly obvious in retrospect: the graviton
needn't live in the same spacetime as the QFT.

2) Hint number two comes from the Holographic Principle 
(a good reference is \cite{Susskind:2005js}).  This is a
far-reaching consequence of black hole thermodynamics.
The basic fact is that a black hole must be assigned an entropy
proportional to the {\it area} of its horizon (in Planck units).
On the other hand, dense matter will collapse into a black hole.
The combination of these two observations leads to the following crazy thing:
The maximum entropy in a region of space is the area of its boundary, in Planck units.
To see this, suppose you have in a volume $V$ (bounded by an area $A$)
a configuration with entropy $S > S_{BH} = {A \over 4 G_N}$
(where $ S_{BH}$ is the entropy of the biggest black hole
fittable in $V$),
but which has {\it less} energy.
Then by throwing in more stuff
(as arbitrarily non-adiabatically as necessary,
\ie\ you can increase the entropy),
since stuff that carries entropy also carries energy\footnote{
Matthew Fisher raises the point that there are systems 
(ones with topological order)
where it is possible to create an information-carrying excitation which 
doesn't change the energy.  I'm not sure exactly how to defend
Bekenstein's argument from this.  I think an important point  
must be that the effects of such excitations are not completely local
(which is why they would be good for quantum computing).
A related issue about which more work has been done is the 
{\it species problem}: if there are many species of fields in the bulk, 
information can be carried by the species label, without any cost in energy.
There are two points which save Bekenstein from this: 
(1) if there are a large number of species of fields, their fluctuations renormalize
the Newton constant (to make gravity weaker), and weaken the bound. 
(2) Being able to encode information in the species label
implies that there is some continuous global symmetry.  It is believed that theories
of quantum gravity do not have continuous global symmetries
(roughly because virtual black holes can eat the charge and therefore generate symmetry-breaking
operators in the effective action, see \eg\ page 12 of \cite{Bigatti:1999dp}).
},
you can {\it make} a black hole.
This would violate the second law of thermodynamics,
and you can use it to save the planet from the humans.
This probably means you can't do it,
and instead we conclude that the black hole
is the most entropic configuration of the theory in this volume.
But its entropy goes like the {\it area}!
This is much smaller than the entropy of a local quantum
field theory on the same space, even with some UV cutoff,
which would have a number of states $N_s \sim e^{ V}$
(maximum entropy $= \ln N_s $).
Indeed it is smaller (when the linear dimensions are large compared to the Planck length) than that of any system with local degrees of freedom,
such as a bunch of spins on a spacetime lattice.

We conclude from this that a quantum theory of gravity
must have a number of degrees of freedom
which scales like that of a QFT in a smaller number of dimensions.
This crazy thing is actually true, and the AdS/CFT 
correspondence 
\cite{Maldacena:1997re}
is a precise implementation of it.

Actually, we already know some examples like this in low dimensions.
An alternative, more general, definition of a quantum gravity is 
a quantum theory where we don't need to introduce the geometry of spacetime (\ie\ the metric)
as input.
We know two ways to accomplish this:\\
a) Integrate over all metrics (fixing some asymptotic data).  This is how GR works.\\
b) Don't ever introduce a metric.  Such a thing is generally called a topological field theory.  The best-understood
example is Chern-Simons gauge theory in three dimensions, where the dynamical variable is a 
one-form field
and the action is $$S_{CS} \sim \int_M \tr A \wedge dA + ... $$
(where the dots is extra stuff to make the nonabelian case gauge invariant);
note that there's no metric anywhere here.
With option (b) there are no local degrees of freedom.
But if you
put the theory on a space with boundary, there are local degrees of freedom which live
on the boundary.  Chern-Simons theory on some three-manifold $M$
induces a WZW model (a 2d CFT) on the boundary of $M$.
So this can be considered an example of the correspondence,
but the examples to be discussed below are quite a bit more dramatic,
because there will be dynamics in the bulk.

3) A beautiful hint as to the possible identity of the extra dimensions is this.
Wilson taught us that a QFT is best thought of as being sliced up by length (or energy) scale,
as a family of trajectories of the renormalization group (RG).
A remarkable fact about this is that the RG equations for the
behavior of the coupling constants as a function of RG scale $u$
are {\it local} in scale:
$$ u \del_u g = \beta(g(u) ) ~.$$
The beta function is determined by the coupling constant evaluated at the energy scale $u$,
and we don't need to know its behavior in the deep UV or IR to figure out how it's changing.
This fact is basically a consequence of locality in ordinary spacetime.
This opens the possibility that we can associate the extra dimensions
suggested by the Holographic idea
with energy scale.
This notion of locality in the extra dimension actually turns out 
to be much {\it weaker} than what we will find in AdS/CFT
(as discussed recently in \cite{Heemskerk:2009pn}),
but it is a good hint.

To summarize, we have three hints for interpreting the Bold Assertion:
\begin{enumerate}
\item The Weinberg-Witten theorem suggests that the graviton lives on a different space than the QFT in question.
\item The holographic principle says 
that the theory of gravity should have a number of degrees of freedom that grows more slowly than the volume.
This suggests that the quantum gravity should live in more  dimensions than the QFT.
\item The structure of the Renormalization Group suggests that we can identify one of these extra dimensions as the RG-scale.
\end{enumerate}

Clearly the field theory in question needs to be strongly coupled.
Otherwise, we can compute and we can see that there is no large 
extra dimension sticking out.  
This is an example of the extremely useful Principle of \\
{\bf Conservation of Evil:} Different weakly-coupled 
descriptions should have non-overlapping regimes of validity.\footnote{
The criterion `different' may require some effort to check.
This Principle is sometimes also called `Conservation of Difficulty'.}

\begin{figure}\begin{center}
\includegraphics[height=130pt]{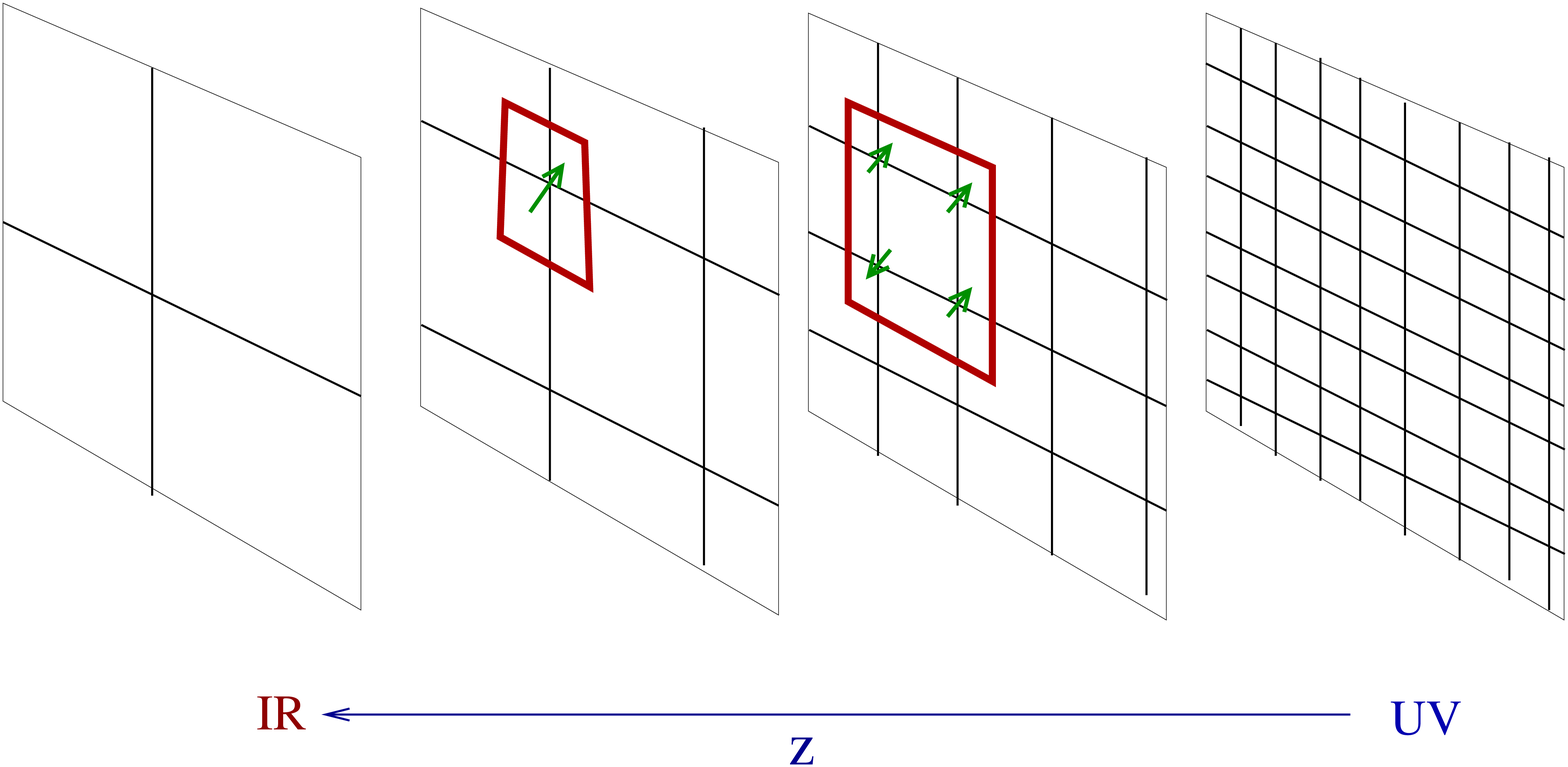} 
\hskip0.5in
\includegraphics[height=130pt]{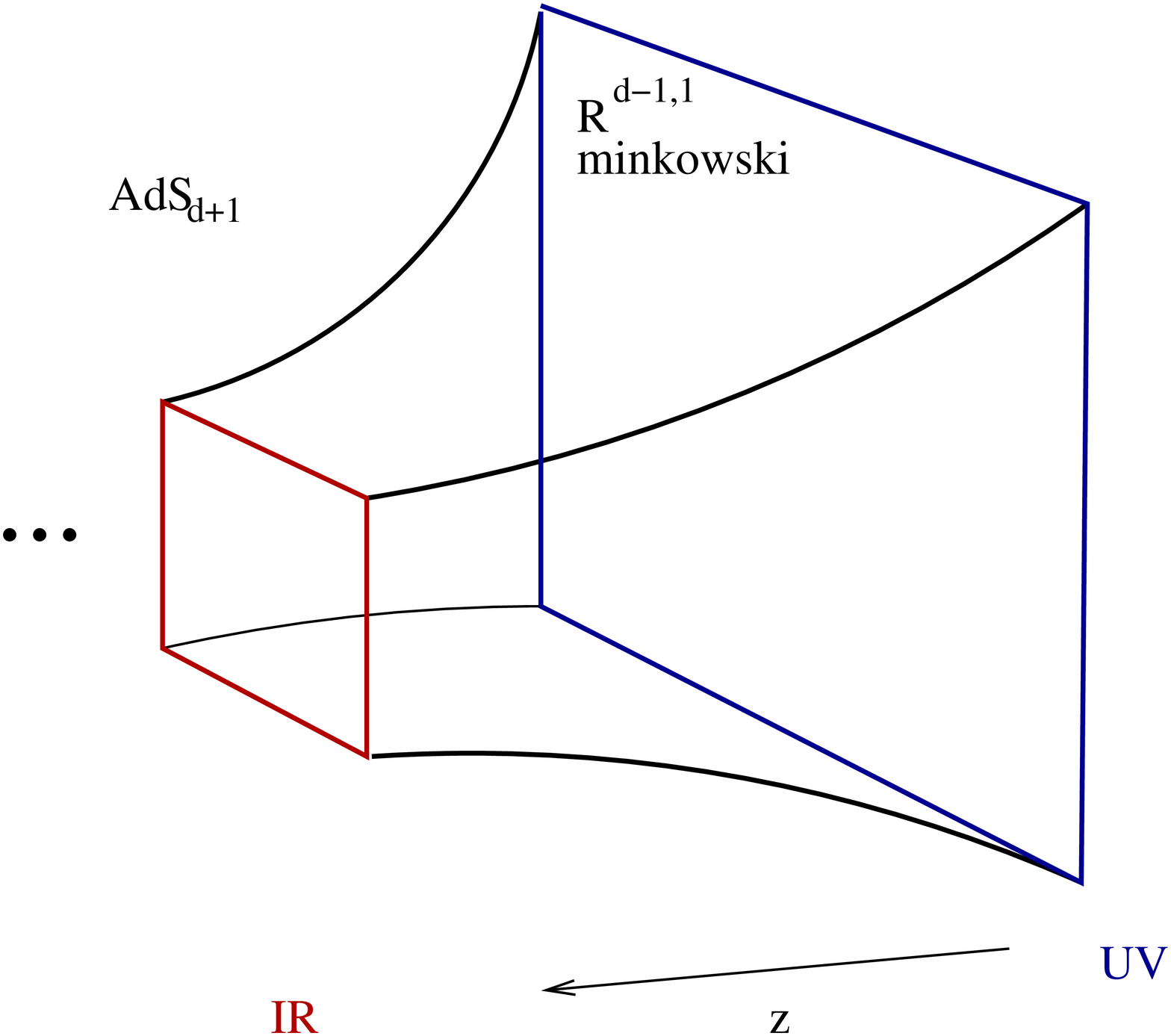} 
\caption{\label{varyingsize}
The extra (`radial') dimension of the bulk is the resolution scale of the field theory.
The left figure indicates a series of block spin transformations labelled by a parameter $z$.
The right figure is a cartoon of AdS space, which organizes the field theory information
in the same way.
In this sense, the bulk picture is a hologram:
excitations with different wavelengths get put in different places in the bulk image.
The connection between these two pictures is pursued further in 
\cite{Swingle:2009bg}.
This paper contains a useful discussion of many features of the correspondence for those familiar with the real-space RG techniques 
developed recently from quantum information theory.
}
\end{center}\end{figure}

Next we will make a simplifying assumption
in an effort to find concrete examples.
The simplest case of an RG flow is when $\beta =0$
and the system is self-similar.  
In a Lorentz invariant theory (which we also assume for simplicity),
this means that the following scale transformation
$x^\mu \to \lambda x^\mu$ ($\mu=0,1, 2, ... d-1$)
is a symmetry.
If the extra dimension coordinate $u$ is to be thought of as an energy scale, then dimensional analysis 
says that $u$ will scale under the scale transformation as $u \to {u \over \lambda}$.
The most general ($d+1$)-dimensional metric (one extra dimension) with this symmetry and Poincar\'e invariance is of the following form:
$$ ds^2 = \({\tilde u \over \tilde L}\) ^2  \eta _{\mu\nu} dx^\mu dx^\nu     + { { d \tilde u ^2  } \over \tilde u^2  }L^2 ~~.$$
We can bring it into a more familiar form by a change of coordinates, 
$\tilde u = {\tilde L \over L} \tilde u$:
$$ ds^2 = \({u \over  L}\) ^2  \eta _{\mu\nu} dx^\mu dx^\nu     + { { d  u ^2  } \over  u ^2  }L^2 
~~.$$ 
This is $AdS_{d+1}$\footnote{It turns out that this metric also has conformal invariance.
So scale and Poincar\'e symmetry implies conformal invariance,
at least when there is a gravity dual.  This is believed to be true more generally \cite{Polchinski:1987dy}, but there is no proof for $d>1+1$.
Without Poincar\'e invariance, scale invariance definitely does {\it not} imply conformal invariance;  indeed there are scale-invariant metrics
without Poincar\'e symmetry, which do not have have special conformal symmetry 
\cite{Kachru:2008yh}.}.  It is a family of copies of Minkowski space,
parametrized by $u$, whose size varies with $u$ 
(see Fig.~\ref{varyingsize}). 
The parameter $L$ is called the `$AdS$ radius' and it has dimensions of length.
Although this is a dimensionful parameter, a scale transformation $x^\mu \to \lambda x^\mu$
can be absorbed by rescaling the radial coordinate $u \to u/\lambda$ (by design);
we will see below more explicitly how this is consistent with scale invariance
of the dual theory.
It is convenient to do one more change of coordinates, to
$ z \equiv {L^2 \over u}$, in which the metric takes the form
\be \label{AdSmetric} ds^2 = \({L \over  z}\) ^2  \( \eta _{\mu\nu} dx^\mu dx^\nu  + dz^2 \)  
~~~.\ee
These coordinates are better because fewer symbols are required to write the metric.
$z$ will map to the length scale in the dual theory.

So it seems that a $d$-dimensional conformal field theory (CFT) should be related to a theory of gravity on $AdS_{d+1}$.
This metric (\ref{AdSmetric}) solves the equations of motion of the following action 
(and many others)\footnote{For verifying statements like this, it can be
helpful to use
Mathematica or some such thing.}
\be \label{bulkaction}
S_{{\rm bulk}}[g, \dots] = {1\over 16 \pi G_N} \int d^{d+1}x \sqrt g \left( - 2 \Lambda + \CR + \dots \right) 
~~.
\ee
Here, $ \sqrt{g} \equiv \sqrt{|\det g|} $ makes the integral
coordinate-invariant, and $\CR$ is the Ricci scalar curvature.
The cosmological constant $\Lambda$ is related by the equations of motion
\be
0 = {\delta S_{{\rm bulk} }\over \delta g^{AB} }
~~\Longrightarrow~~ R_{AB} + {d  \over L^2} g_{AB} = 0 
\ee
to the value of the AdS radius: 
$ - 2 \Lambda = { d(d-1) \over L^2} $.
This form of the action (\ref{bulkaction}) is what we would guess using Wilsonian 
naturalness (which in some circles is called the `Landau-Ginzburg-Wilson paradigm'):
we include all the terms which respect the symmetries (in this case, this is
general coordinate invariance), organized by decreasing relevantness, \ie\ by the number of derivatives.  
The Einstein-Hilbert term (the one with the Ricci scalar) is an irrelevant operator:
$ \CR \sim \partial^2 g + (\partial g)^2 $ has dimensions of length${}^{-2}$, 
so $G_N $ here is a length${}^{d-1}$, the Planck length: $G_N \equiv \ell_{pl}^{d-1} \equiv
M_{pl}^{1-d}$ (in units where $\hbar = c=1$).
The gravity theory is classical if $ L \gg \ell_{pl}$.
In this spirit, the $\dots$ on the RHS denote more irrelevant terms involving
more powers of the curvature.
Also hidden in the $\dots$ are other bulk fields which vanish in the dual of the CFT vacuum 
(\ie\ in the AdS solution).

This form of the action (\ref{bulkaction}) is indeed what comes from string theory at 
low energies and when the curvature (here, $\CR \sim {1\over L^2}$) is small 
(compared to the {\it string tension}, $ {1\over \alpha'} \equiv {1\over \ell_s^2} $;
this is the energy scale that determines the masses of excited vibrational modes of the
string),
at least in cases where we are able to tell.
The main role of string theory in this business 
(at the moment) is to provide consistent ways 
of filling in the dots.

In a theory of gravity, the space-time metric is a dynamical variable, 
and we only get to specify the boundary behavior.
The $AdS$  metric above has a boundary at $z=0$.
This is a bit subtle.
Keeping $x^\mu$ fixed and moving in the $z$ direction
from a finite value of $z$ to $z=0$ is actually infinite distance.
However, massless particles in $AdS$ (such as the 
graviton discussed above) travel along null geodesics;
these reach the boundary in finite time.
This means that in order to specify the future evolution
of the system from some initial data,
we have also to specify boundary conditions at $z=0$.
These boundary conditions will play a crucial role in the discussion below.

So we should amend our statement to say that 
a $d$-dimensional conformal field theory is related to a theory of gravity on
spaces which are {\it asymptotically} $AdS_{d+1}$.
Note that this case of negative cosmological constant (CC)
turns out to be much easier to understand holographically than the 
naively-simpler (asymptotically-flat) case of zero CC.  Let's not even talk about the case of 
positive CC (asymptotically de Sitter).

Different CFTs will correspond to such theories of gravity 
with different field content and different bulk actions, \eg\ different 
values of the coupling constants in $S_{{\rm bulk}}$.
The example which is understood best is the case of the $\CN = 4$ 
super Yang-Mills theory (SYM) in four dimensions.
This is dual to maximal supergravity in $AdS_5$
(which arises by dimensional reduction of ten-dimensional IIB 
supergravity on $AdS_5 \times S^5$).
In that case, we know the precise values of many of the coefficients in the bulk action.
This will not be very relevant for our discussion below.
An important conceptual point is that the values of the bulk parameters 
which are realizable will in general be discrete\footnote{An example of this
is the relationship (\ref{speciesrelation}) between the Newton constant in the bulk 
and the {\it number} of species in the field theory, which we will find in the next subsection.}.
This discreteness is hidden by the classical limit.

We will focus on the case of relativistic CFT for 
a while, but let me emphasize here that the name `AdS/CFT' 
is a very poor one: the correspondence is much more general.
It can describe deformations of UV fixed points by relevant operators,
and it has been extended to cases which are 
not even relativistic CFTs in the UV:
examples include fixed points with dynamical critical
exponent $z \neq 1$ \cite{Kachru:2008yh}, 
Galilean-invariant theories \cite{Son:2008ye, Balasubramanian:2008dm},
and theories which do more exotic things in the UV like the `duality cascade' of
\cite{Klebanov:2000hb}.

\subsection{Counting of degrees of freedom}\label{sec:counting}

We can already make a check of the conjecture that a gravity theory
in $AdS_{d+1}$ might be dual to a QFT in $d$ dimensions.
The holographic principle tells us that the area of the boundary in Planck units is the number of degrees of freedom (dof), \ie\ the maximum entropy:
$${\rm Area~of~boundary  \over 4G_N} ~{\buildrel{?}\over{ =}} ~
{\rm number~of~dof~of~QFT }\equiv N_d ~~.$$
Is this true \cite{Susskind:1998dq}? 
Yes: both sides are equal to infinity. We need to regulate our counting.

Let's regulate the field theory first. 
There are both UV and IR divergences.
We put the thing on a lattice, introducing a 
short-distance cut-off $\epsilon$ 
(\eg, the lattice spacing) and we put it in a cubical box of linear size $R$. The total number of degrees of freedom is the number of cells $ \left( {R \over \epsilon } \right)^{d-1} $, times 
the number of degrees of freedom per lattice site, which we will call `$\hl{N^2}$'.
The behavior suggested by the name we have given this number 
is found in well-understood examples.
It is, however, clear (for example from the structure of known $AdS$ vacua of string theory \cite{Silverstein:2003jp}) that other behaviors $N^b$ are possible, and that's why I made it a funny color and put it in quotes.
So $N_d={R^{d-1} \over \epsilon^{d-1}}\hl{N^2}$.

The picture we have of AdS${}_{d+1}$ is a collection of copies of 
$d$-dimensional Minkowski space of varying size; the boundary is the locus $z \to 0$ where they get really big. The area of the boundary is
\be A=\int_{\IR^{d-1}, ~z \to 0,~{\rm fixed}~t} \sqrt g d^{d-1}x = \int_{\IR^{d-1}, ~z \to 0} d^{d-1}x {L^{d-1} \over z^{d-1}} ~~.\ee
As in the field theory counting, this is infinite for two reasons: from the integral over $x$ and from the fact that $z$ is going to zero.
To regulate this integral, we integrate not to $z=0$ but rather cut it off at $z=\epsilon$.  We will see below a great deal more evidence for this idea that the boundary of $AdS$ is associated
with the UV behavior of the field theory,
and that cutting off the geometry at $z=\epsilon$ is a UV cutoff (not identical 
to the lattice cutoff, but close enough for our present purposes).
Given this,
\be A =\int_0^R d^{d-1}x{L^{d-1} \over z^{d-1}} |_{z=\epsilon}  =  \left({R L\over \epsilon}\right)^{d-1} ~~.\ee
The holographic principle then says that the maximum entropy in the bulk is 
\be {A \over 4 G_N} \sim  {L^{d-1} \over 4 G_N}  \left({R \over \epsilon}\right)^{d-1} .\ee

We see that the scaling with the system size agrees -- the both-hand-side goes like
$R^{d-1}$.
So AdS/CFT is indeed an implementation of the holographic principle.
We can learn more from this calcluation: 
In order for the prefactors of $R^{d-1}$ to agree, 
we need to relate the 
$AdS$ radius in Planck units 
$ {L^{d-1} \over G_N} \sim (L M_{pl})^{d-1}$
to the number of degrees of freedom per site
of the field theory:
\be\label{speciesrelation}
 \boxed{{L^{d-1} \over G_N}  = \hl{N^2} } ~\ee
up to numerical prefactors.

\subsection{Preview of the AdS/CFT correspondence}

Here's the ideology:
\begin{eqnarray*}
\textrm{fields in AdS} & \longleftrightarrow & \textrm{local operators of CFT} \\
\textrm{{\small spin}} & & \textrm{{\small spin}}\\
\textrm{{\small mass}} & & \textrm{{\small scaling dimension $\Delta$}}
\end{eqnarray*}
In particular, for a scalar field in AdS, 
the formula relating the mass of the scalar field 
to the scaling dimension of the corresponding operator in the CFT is $m^2 L_{AdS}^2 = \Delta (\Delta - d)$, as we'll show in section \ref{sec:refinement}.

One immediate lesson from this formula is that 
a simple bulk theory with a small number of light fields
is dual to a CFT with a hierarchy in its spectrum of operator dimensions.
In particular, there need to be a small number of operators with small 
(\eg\ of order $\hl{N}^0$) dimensions.  If you are aware of explicit examples of such theories,
please let me know\footnote{Rational CFTs in two dimensions don't count because they 
fail our other criterion for a simple gravity dual: 
in the case of a 2d CFT, the central charge of the Virasoro algebra, $c$, is 
a good measure of $`\hl{N^2}'$, the number of degrees of 
freedom per point. But rational CFTs have $c$ of order unity, and therefore 
can only be dual to very quantum mechanical theories of gravity.
But this is the right idea. 
Joe Polchinski has referred to the
general strategy being applied here
as ``the Bootstrap for condensed matter physics".  
The connection with the 
bootstrap in its CFT incarnation 
\cite{Belavin:1984vu}
is made quite direct in \cite{Heemskerk:2009pn}.
}\footnote{
Eva Silverstein and Shamit Kachru have emphasized that 
this special property of these field theories is 
a version of the `cosmological constant problem,' \ie\ it is
dual to the 
specialness of having a small cosmological constant in the bulk.  
At least in the absence of supersymmetry, 
there is some
tuning that needs to be done in the landscape of string vacua to choose
these vacua with a small vacuum energy, and hence a large AdS radius.
Here is a joke about this: 
when experimentalists look at some material
and see lots of complicated crossovers, they will tend to throw it away;
if they see instead some simple beautiful power laws, as would
happen in a system with few low-dimension operators, they will keep it.  
Perhaps these selection effects are dual to each other.
}.
This is to be distinguished from the thus-far-intractable case where some whole
tower of massive string modes in the bulk are needed.

Now let's consider some observables of a QFT (we'll assume Euclidean spacetime for now),
namely vacuum correlation functions of local operators in the CFT:
\begin{equation*}
\vev{\CO_1(x_1) \CO_2(x_2) \cdots \CO_n(x_n)}~~.
\end{equation*}
We can write down a generating functional $Z[J]$ for these correlators by
perturbing the action of the QFT:
\begin{gather*}
\CL(x) \rightarrow \CL(x) + \sum_A J_A(x) \CO_A(x) \equiv \CL(x) + \CL_J(x)\\
Z[J] = \vev{e^{-\int \CL_J}}_{CFT}~~~.
\end{gather*}
where $J_A(x)$ are arbitrary functions (sources) and $\{\CO_A(x)\}$ is some basis of local operators.
The $n$-point function is then given by:
\begin{equation*}
\vev{\prod_n \CO_n(x_n)} = \prod_n \frac{\delta}{\delta J_n(x_n)} \ln Z\Big|_{J=0}~~.
\end{equation*}

Since $\CL_J$ is a \emph{UV} perturbation (because it is a perturbation of the \emph{bare} Lagrangian by \emph{local} operators), in AdS it corresponds to a perturbation near the boundary, $z \to 0$. 
(Recall from the counting of degrees of freedom in 
section \ref{sec:counting} that QFT with UV cutoff $E < 1/\epsilon$ $\longleftrightarrow$ AdS cutoff $z > \epsilon$.)
The perturbation $J$ of the CFT action will be encoded in the boundary condition on bulk fields.

%

The idea (\cite{Gubser:1998bc, Witten:1998qj}, often referred to as GKPW) for computing $Z[J]$ is then, schematically:
\begin{equation}\label{GKPW}
Z[J] \equiv \vev{e^{-\int \CL_J}}_{CFT} = \underbrace{Z_{\textrm{QG}}[\textrm{b.c.\@ depends on $J$}]}_{=???}
\underset{\hl{N} \gg 1}{\sim} 
	e^{-S_{\textrm{grav}}}\big|_{\textrm{EOM, b.c.\@ depend on $J$}}~~.
\end{equation}
The middle object is the partition function of quantum gravity.
We don't have a very useful idea of what this is, except in perturbation theory
and via this very equality.  
In a limit where this gravity theory becomes classical, however,
we know quite well what we're doing, and we can do the path integral by saddle point,
as indicated on the RHS of (\ref{GKPW}).

An important point here is that even though we are claiming that
the QFT path integral is dominated by a classical saddle point,
this does not mean that the field theory degrees of freedom are free.
How this works depends on what kind of large-$N$ limit we take
to make the gravity theory classical.  This is our next subject.


\section{When is the gravity theory classical?}

So we've said that some QFT path integrals are dominated
by saddle points\footnote{Note that I am not saying here that 
the configuration of the elementary fields in the path integral necessarily 
have some simple description at the saddle point.  Thanks to Larry Yaffe for 
emphasizing this point.} where the degrees of freedom near the saddle 
are those of a gravitational theory in extra dimensions:
\be
Z_{some~QFTs}[{\rm sources}] \approx e^{ - S_{{\rm bulk}}[{\rm boundary~conditions~at~}z\to 0]}|_{{\rm extremum~of~}S_{{\rm bulk}}} .
\ee
The sharpness of the saddle (the size of the second derivatives of the action evaluated
at the saddle)
is equivalent to the classicalness of the bulk theory.
In a theory of gravity,
this is controlled by the Newton constant in front of the action.
More precisely,
in an asymptotically-$AdS$ space with $AdS$ radius $L$,
the theory is classical when
\be {L^{d-1} \over G_N } \equiv `\hl{N^2}{}' \gg 1 .\ee
This quantity, the $AdS$ radius in Planck units 
${L^{d-1} \over G_N } \equiv \(L M_{pl}\)^{d-1}$, 
is what we identified (using the holographic principle) 
as the number of degrees of freedom
per site of the QFT.  

In the context of our current goal, it is worth spending some 
time talking about different kinds of large-species limits of QFTs.
In particular, in the condensed matter literature, the phrase `large-enn' 
usually means that one promotes a two-component object
to an $n$-component vector, with $O(n)$-invariant interactions.
This is probably {\it not} what we need to have a simple gravity
dual, for the reasons described next.

\subsection{Large $n$ vector models}

A simple paradigmatic example of this vector-like large-$n$ limit
(I use a different $n$ to distinguish it from the matrix case to be discussed next) is 
a QFT of $n$ scalar fields $\vec \varphi = (\varphi_1, \dots, \varphi_n)$ 
with the following action 
\be
S[\varphi] = - \half \int d^dx \left( \partial_\mu \vec \varphi \partial^\mu\vec \varphi 
+ m^2 \vec \varphi \cdot \vec \varphi + { \lambda_v \over n} \(\vec \varphi \cdot \vec \varphi\)^2 \right) ~~.
\ee
The fields $\vec \varphi$ transform in the fundamental representation of 
the $O(n)$ symmetry group.  
Some foresight has been used to determine that the quartic coupling $\lambda_v$ is to be held fixed in the large-$n$ limit.
An effective description (\ie\ a well-defined saddle-point) 
can be found in terms of 
$ \sigma \equiv \vec \varphi \cdot \vec \varphi$ by standard
path-integral tricks, and the effective action for $\sigma$ is
\be S_{{\rm eff}}[\sigma] = - {n\over 2} \left[ \int { \sigma^2 \over 2 \lambda}  + \tr \ln\(- \partial^2 + m^2 + \sigma \)  \right].\ee
The important thing is the giant factor of $n$ in front of the action 
which makes the theory of $\sigma$ classical.  
Alternatively, the only interactions in this $n$ vector model are "cactus" diagrams;
this means that, modulo some self energy corrections, the theory is free.

So we've found a description of this saddle point {\it within} weakly-coupled 
quantum field theory.  
The Principle of Conservation of Evil then suggests that this should not 
also be a simple, classical theory of gravity.
Klebanov and Polyakov 
\cite{Klebanov:2002ja} have suggested what the (not simple) gravity dual might be.

%

\def\gYM{g}
\subsection{'t Hooft counting}

\begin{flushright}
{\it ``You can hide a lot in a large-$N$ matrix."}\\
 -- Steve Shenker \end{flushright}

Given some system with a few degrees of freedom,
there exist many interesting large-$N$ generalizations,
many of which may admit saddle-point descriptions.
It is not guaranteed that  the effective degrees of freedom near the saddle (sometimes ominously called `the 
masterfield')
 are simple field theory degrees of freedom
(at least not in the same number of dimensions).
If they are not, this means that such a limit isn't immediately useful,
but it isn't necessarily more distant from the 
physical situation than the limit of the previous subsection.  
In fact, we will see dramatically below that 
the 't Hooft limit described here preserves {\it more} features
of the interacting small-$N$ theory than the usual vector-like limit.
The remaining problem is to find a description of the masterfield, and this is
precisely what's accomplished by AdS/CFT.

Next we describe in detail a large-$N$ limit (found by 't Hooft\footnote{The standard
pedagogical source for this material is \cite{Coleman}, available 
  \htmladdnormallink{from the 
  KEK KISS server.}{http://ccdb4fs.kek.jp/cgi-bin/img_index?8005163}}) where the 
right degrees of freedom seem to be closed strings
(and hence gravity).
In this case, the number of degrees of freedom per point in the QFT will go like
$N^2$.
Evidence from the space of string vacua 
suggests that there are many generalizations of this 
where the number of dofs per point goes like $N^b$ for $b \neq 2$ \cite{Silverstein:2003jp}.  
However, a generalization of the 't Hooft limit is not yet well-understood 
for other cases\footnote{Recently, there has been an explosion 
of literature on a case where the number of degrees of freedom per point 
should go like $N^{3/2}$ \cite{ABJM}.}.

Consider a (any) quantum field theory with \emph{matrix} fields, $\Phi_{a=1,\ldots,\bar{N}}^{b=1,\ldots,N}$.
By matrix fields, we mean that their products appear in the Lagrangian only in the form of matrix multiplication, \eg\ $(\Phi^2)_a^c = \Phi_a^b \Phi_b^c$,
which is a big restriction on the interactions.  
It means the interactions must be invariant under 
$\Phi \to U^{-1} \Phi U$;
for concreteness we'll take the matrix group to be $U \in U(N)$\footnote{
Note that the important distinction between these models and those 
of the previous subsection is not the difference in groups ($U(N)$ vs $O(N)$), but rather the
difference in representation in which the fields transform: here the fields 
transform in the adjoint representation rather than the fundamental.}.
The fact that this theory has many more interaction terms than the vector
model with the same number of fields (which would have a much larger $O(N^2)$ symmetry) 
changes the scaling of the coupling in the large $N$ limit.

In particular, consider the \emph{'t Hooft limit} in which $N \rightarrow \infty$ and $\gYM \rightarrow 0$ with $\lambda = \gYM^2 N$ held fixed in the limit.  Is 
the theory free in this limit? The answer turns out to be no. The loophole is that even though the coupling goes to zero, the number of modes diverges.
Compared to the vector model, 
the quartic coupling in the 
matrix model $\gYM \sim 1/\sqrt{N}$
goes to zero slower than the coupling in 
the 
vector model $g_v \equiv \lambda_v/N \sim 1/N$.  

We will be agnostic here about whether the $U(N)$ symmetry is gauged,
but if it is not there are many more states than we can handle using the 
gravity dual.  The important role of the gauge symmetry for our purpose
is to restrict the physical spectrum to 
gauge-invariant operators, like $\tr \Phi^k$.

The fields can have all kinds of spin labels and global symmetry labels,
but we will just call them $\Phi$.  In fact, the location in space can also 
for the purposes of the discussion of this section be considered as merely a label on the field 
(which we are suppressing).  
So consider a schematic Lagrangian of the form:
\begin{equation*}
\CL \sim \frac{1}{\gYM^2} \Tr\left((\p \Phi)^2 + \Phi^2 + \Phi^3 + \Phi^4 + \ldots \right)~~.
\end{equation*}
I suppose we want $\Phi$ to be Hermitian so that this Lagrangian is real,
but this will not be important for our considerations.

\begin{figure}[h] 
\begin{center}
	\subfigure[]{\includegraphics[scale=0.6]{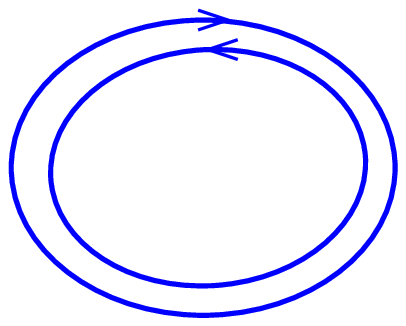} \raisebox{5ex}[0cm][0cm]{$\propto N^2$}} \quad
	\subfigure[]{\includegraphics[scale=0.6]{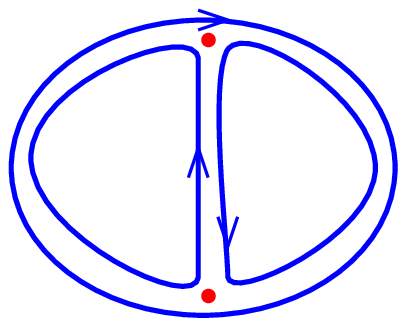} \raisebox{5ex}[0cm][0cm]{$\propto \lambda N^2$}} \quad
	\subfigure[]{\includegraphics[scale=0.5]{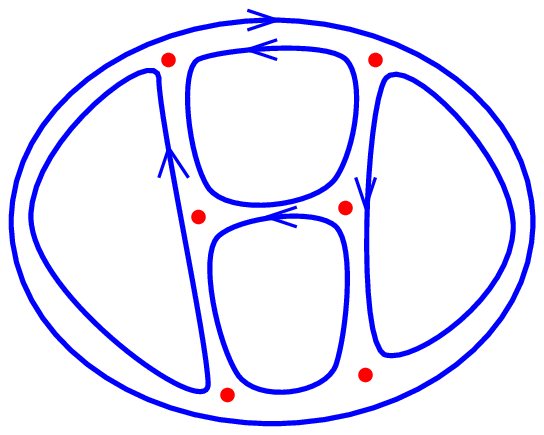} \raisebox{5ex}[0cm][0cm]{$\propto \lambda^3 N^2$}} \quad
	\caption{\label{fig:vac_planer} planar graphs that contribute to the vacuum$\rightarrow$vacuum amplitude.}
\end{center} \end{figure}

\begin{figure}[h] \begin{center}
	\includegraphics[scale=0.5]{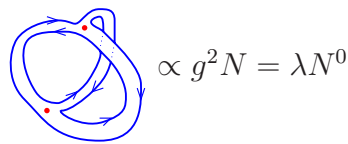} \raisebox{5ex}[0cm][0cm]{$\propto \gYM^2 N = \lambda N^0$}
	\caption{\label{fig:vac_nonplaner} Non-planar (but still oriented!) graph that contributes to the vacuum$\rightarrow$vacuum amplitude.
	}
\end{center} 
\end{figure}

We will now draw some diagrams which let us keep track of the 
$N$-dependence of various quantities.  It is convenient to adopt the \emph{double line} notation, in which oriented index lines follow conserved color flow.   
We denote the propagator by\footnote{Had we been considering $SU(N)$, the result would be $\vev{\Phi_b^a \Phi_c^d} \propto \delta_c^a \delta_b^d - \delta_b^a \delta_d^c/N^2$ = \raisebox{-1ex}[0cm][0cm]{\includegraphics[scale=0.8]{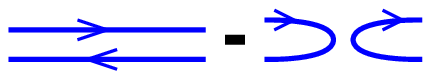}}~.  This difference can be ignored at leading order 
in the $1/N$ expansion.
}:
\begin{equation*}
\vev{ \Phi_b^a \Phi_c^d} \propto \gYM^2 \delta_c^a \delta_b^d \equiv
\gYM^2 ~\raisebox{-1.5ex}[0cm][0cm]{\includegraphics[scale=0.4]{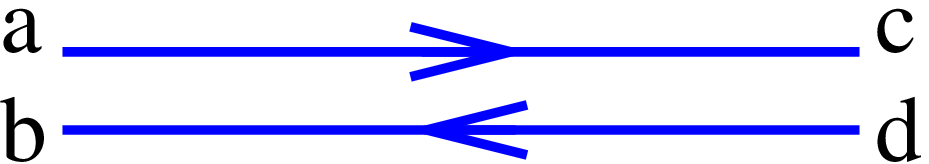}} \\
\end{equation*}
and the vertices by:
\marginpar{
\rotatebox{270}
{\tiny [Brian Swingle]}}
\begin{xalignat*}{2}
\includegraphics[scale=.2]{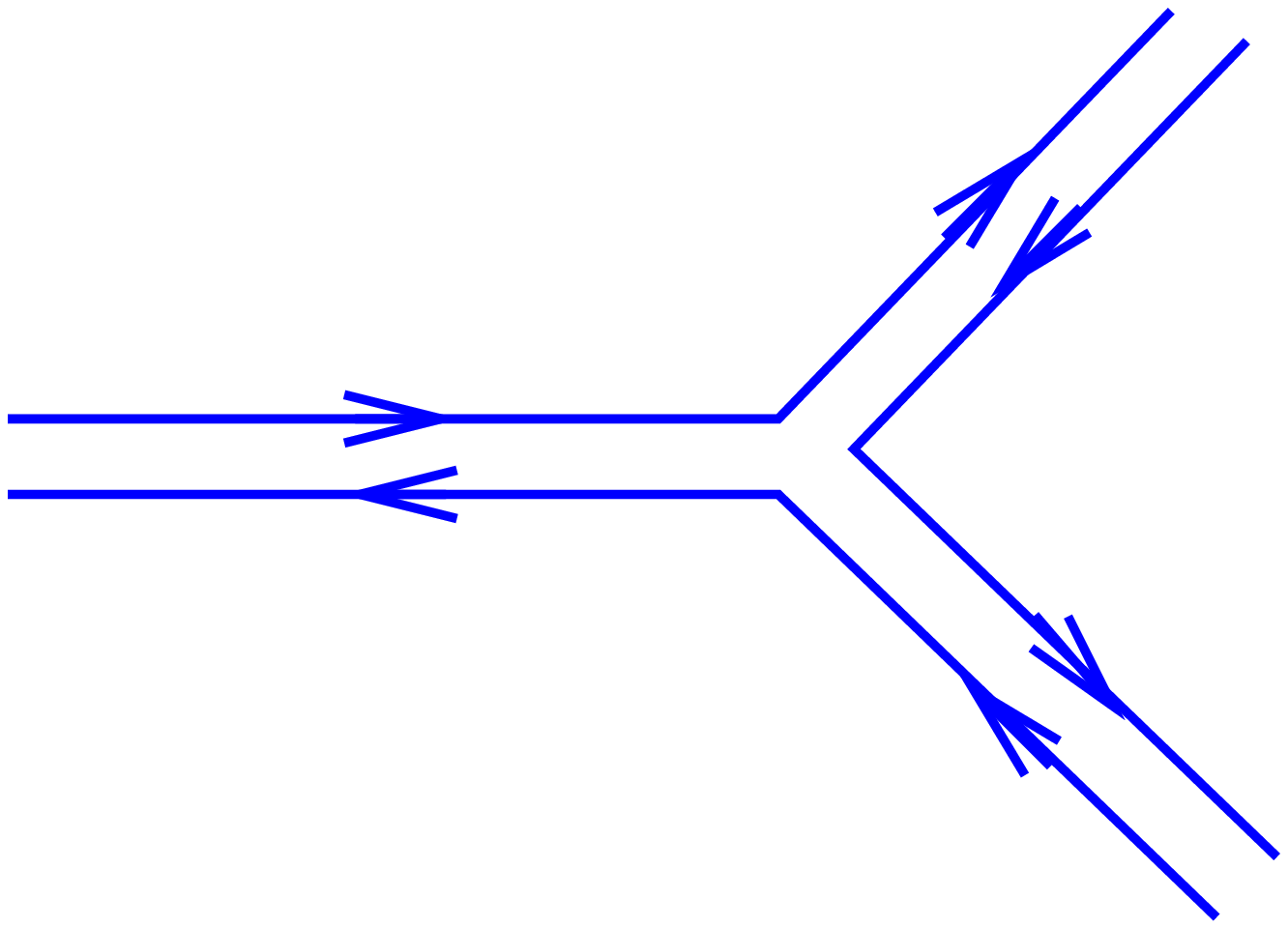} & \raisebox{7ex}[0cm][0cm]{$\propto \gYM^{-2}$} & 
\includegraphics[scale=.2]{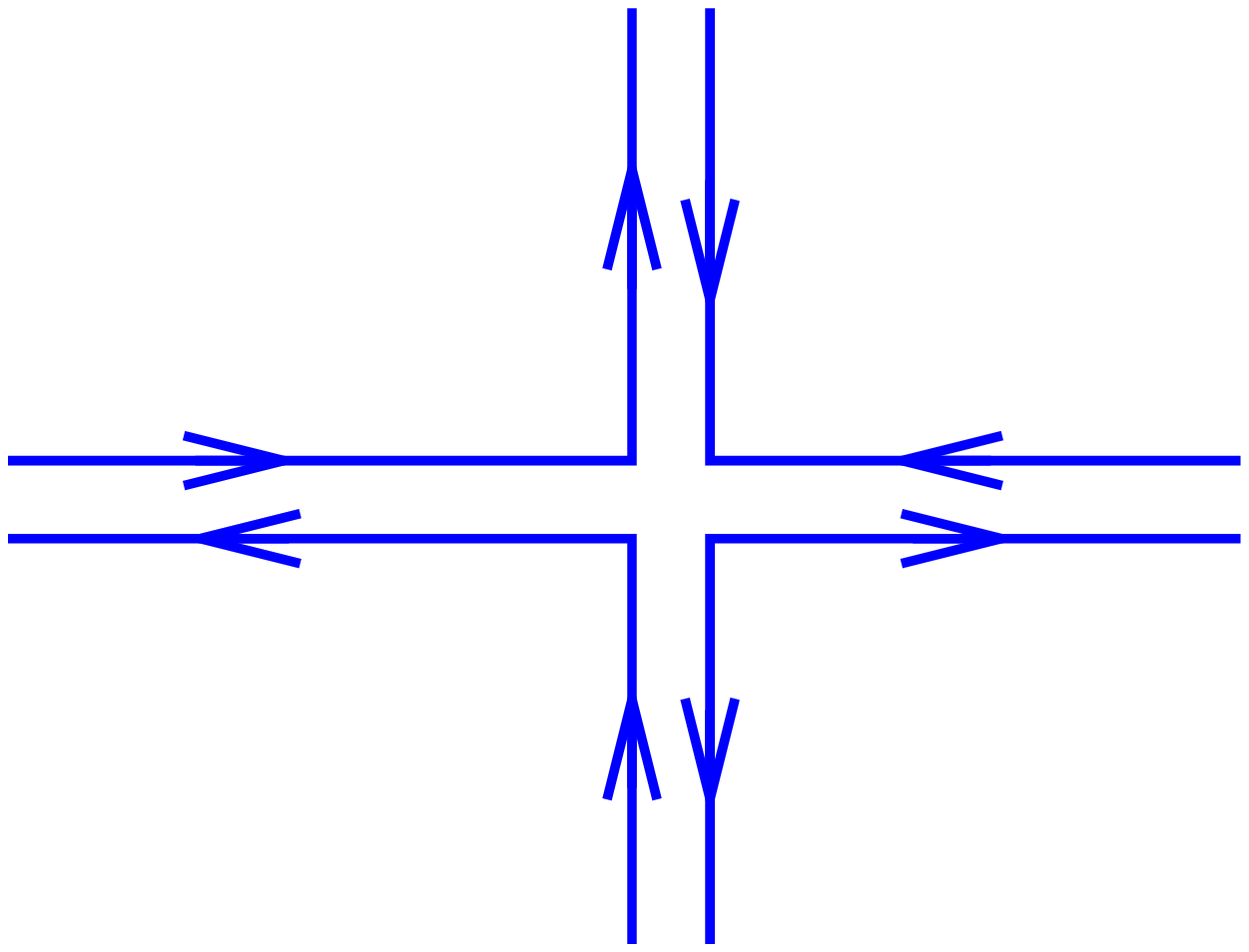} & \raisebox{7ex}[0cm][0cm]{~~$\propto \gYM^{-2}$} ~~.
\end{xalignat*}
To see the consequences of this more concretely, let's consider some vacuum-to-vacuum diagrams (see Fig.~\ref{fig:vac_planer} and \ref{fig:vac_nonplaner} for illustration). 
We will keep track of the color structure, and not worry even about
how many dimensions we are in (the theory could even be zero-dimensional, such as the matrix integral
which constructs the Wigner-Dyson distribution).

\begin{figure}[h]
\begin{center}
\includegraphics{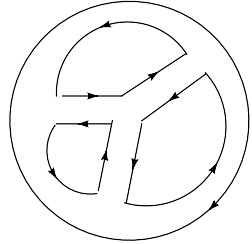}
\caption{\label{example1}This diagram consists of $4$ three point vertices, $6$ propagators, and $4$ index loops}
\end{center}
\end{figure}

A general diagram consists of propagators, interaction vertices, and index loops, and gives a contribution
\begin{equation}
\mbox{diagram} \sim \left(\frac{\lambda}{N}\right)^{\mbox{no. of prop.}} \left(\frac{N}{\lambda}\right)^{\mbox{no. of int. vert.}} N^{\mbox{no. of index loops}}.
\end{equation}
For example, the diagram in Fig.~\ref{example1} has $4$ three point vertices, $6$ propagators, and $4$ index loops, giving the final result $N^2 \lambda^2$.
In Fig.~\ref{fig:vac_planer} we have a set of \emph{planar graphs}, 
meaning that we can draw them on a piece of paper without any lines crossing;
their contributions take the general form $\lambda^n N^2$. However, there also exist \emph{non-planar graphs}, such as the one in Fig.~\ref{fig:vac_nonplaner}, whose contributions are down by (an even number of) powers of $N$.
One thing that's great about this expansion is that the diagrams which
are harder to draw are less important.
\marginpar{
\rotatebox{270}
{\tiny [Wing-Ko Ho]}}

\begin{figure}[h] \begin{center}
	\subfigure[]{\includegraphics[scale=0.6]{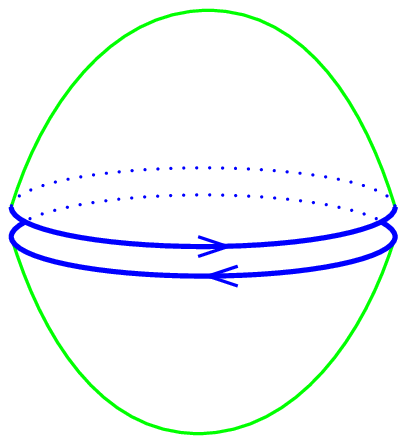} \raisebox{5ex}[0cm][0cm]{$\sim S^2$}} \quad
	\subfigure[]{\includegraphics[scale=0.6]{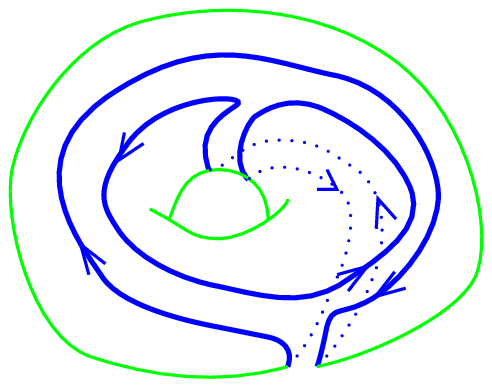} \raisebox{5ex}[0cm][0cm]{$\sim T^2$}}
	\caption{\label{fig:vac_surf} Direct surfaces constructed from the vacuum diagram in (a) Fig.~\ref{fig:vac_planer}a and (b) Fig.~\ref{fig:vac_nonplaner}.}
\end{center} \end{figure}

\begin{figure}[h] \begin{center}
	\includegraphics[scale=0.5]{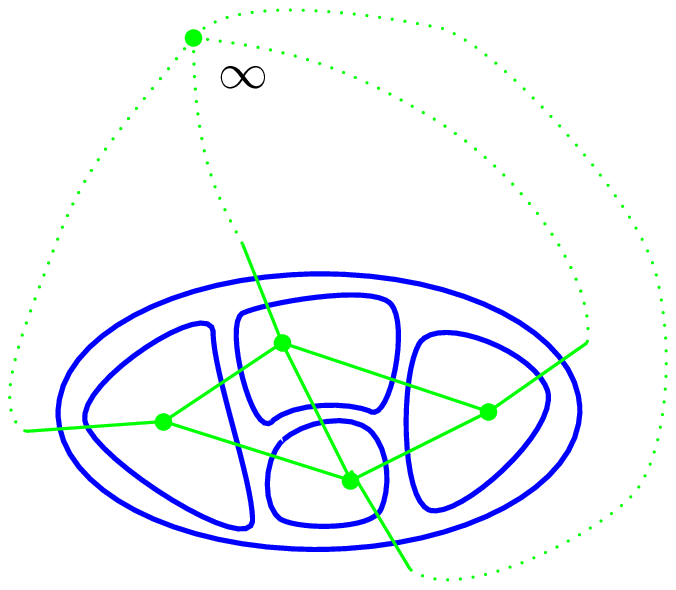} \raisebox{5ex}[0cm][0cm]{$\sim S^2$}
	\caption{\label{fig:vac_surf_dual} Dual surface constructed from the vacuum diagram in Fig.~\ref{fig:vac_planer}c. Note that points at infinity are identified.}
\end{center} \end{figure}

We can be more precise about how the diagrams are organized.
Every double-line graph specifies a triangulation of a 2-dimensional surface $\Sigma$. There are two ways to construct the explicit mapping: 
\begin{description}
\item[Method 1 (``direct surface'')] Fill in index loops with little plaquettes.
\item[Method 2 (``dual surface'')](1) draw a vertex\footnote{Please don't be confused by multiple uses of the word `vertex'.  There are interaction vertices of various kinds in the Feynman diagrams and these correspond to vertices in the triangulation only in the first formulation.}
in every index loop and (2) draw an edge across every propagator.
\end{description}
These constructions are illustrated in Fig.~\ref{fig:vac_surf} and \ref{fig:vac_surf_dual}.


%

If $E=$ number of propagators,
$V=$ number of vertices,
and $F=$ number of index loops,
then the diagram gives a contribution $N^{F-E+V}\lambda^{E-V}$.  The letters refer to 
the `direct' triangulation of the surface in which interaction vertices are triangulation vertices.  Then we interpret $E$ as the number of edges, $F$ as the number of faces, and $V$ as the number of vertices in the triangulation.  In the dual triangulation there are dual faces $\tilde{F}$, dual edges $\tilde{E}$, and dual vertices $\tilde{V}$.  The relationship between the original and dual variables is $E = \tilde{E}$, $V = \tilde{F}$, and $F = \tilde{V}$.  
The exponent $\chi = F - E + V = \tilde{F}-\tilde{E}+\tilde{V}$ is the Euler character and it is a topological invariant of two dimensional surfaces.  In general it is given by $\chi(\Sigma) = 2 - 2h - b$ where $h$ is the number of handles (the genus) and $b$ is the number of boundaries.  Note that the exponent of $\lambda$, $E-V$ or $\tilde{E} - \tilde{F}$ is not a topological invariant and depends on the triangulation (Feynman diagram).


Because the $N$-counting is topological (depending only on $\chi(\Sigma)$) we can sensibly organize the perturbation series for the effective action $\ln Z$ in terms of a sum over surface topology.  Because we're computing only vacuum diagrams for the moment, the surfaces we're considering have no boundaries $b=0$ and are classified by their number of handles $h$ ($h=0$ is the two dimensional sphere, $h=1$ is the torus, and so on).  We may write the effective action (the sum over connected vacuum-to-vacuum diagrams) as
\begin{equation}
\ln{Z} = \sum^\infty_{h=0} N^{2-2h} \sum^\infty_{\ell = 0} c_{\ell, h} \lambda^\ell = \sum^\infty_{h=0} N^{2-2h} \mathfrak{F}_h(\lambda)
\end{equation}
where the sum over topologies is explicit.

Now we can see some similarities between this expansion and perturbative string expansions\footnote{The following two paragraphs may be skipped by the reader
who doesn't want to hear about string theory.}.
$1/N$ plays the role of the string coupling $g_s$, 
the amplitude joining and splitting of the closed strings.
In the large $N$ limit, this process is suppressed and the theory is classical.
Closed string theory generically predicts gravity, with Newton's constant
$G_N \propto g_s^2$, so this reproduces our result $G_N \sim N^{-2}$
from the holographic counting of degrees of freedom (this time, without the quotes around it).

It is reasonable to ask what plays the role of the worldsheet coupling:
there is a 2d QFT living on the worldsheet of the string,
which describes its embeddings into the target space;
this theory has a weak-coupling limit when the target-space curvature $L^{-2}$ is small,
and it can be studied in perturbation theory in powers of $\ell_s \over L$, where $\ell_s^{-2}$
is the string tension.  We can think of $\lambda$ as a sort of chemical potential for edges in our triangulation.  Looking back at our diagram counting we can see that if $\lambda$ becomes large then diagrams with lots of edges are important.  Thus large $\lambda$ encourages a smoother triangulation of the worldsheet which we might interpret as fewer quantum fluctuations on the worldsheet.  We expect a relation of the form $\lambda^{-1} \sim \alpha'$ which encodes our intuition about large $\lambda$ suppressing fluctuations.
This is what is found in well-understood examples.

This story is very general in the sense that all matrix models define something like a theory of two-dimensional fluctuating surfaces via these random triangulations.  The connection is even more interesting when we remember all the extra labels we've been suppressing on our field $\Phi$.  For example, the position labeling where the field $\Phi$ sits plays the role of embedding coordinates on the worldsheet.  Other indices (spin, etc.) indicate further worldsheet degrees of freedom.  
 However, the microscopic details of the worldsheet theory are not so easily discovered.  
 It took about fifteen years between the time when 't Hooft described
 the large-$N$ perturbation series in this way,  
 and the first examples where the worldsheet dynamics were identified 
 (these old examples are reviewed in \eg\ \cite{MatrixQM}).
%

As a final check on the non-triviality of the theory in the 't Hooft limit, let's see if the 't Hooft coupling runs with scale.  For argument let's think about the case when
the matrices are gauge fields and $L =-{1\over g_{YM}^2}  \tr F_{\mu\nu} F^{\mu\nu}$.
In $d$ dimensions, the behavior through one loop is 
\be
\mu \partial_\mu \, g_{YM} \equiv
\beta_g \sim \frac{4-d}{2} g_{YM} + b_0 g_{YM}^3 N ~.\ee
($b_0$ is a coefficient which depends on the matter content, and vanishes
for $\CN=4$ SYM.)
So we find that 
$\beta_\lambda \sim  \frac{4-d}{2} \lambda + b_0 \lambda^2 $.
%
Thus $\lambda$ can still run in the large $N$ limit and the theory is non-trivial.  

\subsection{$N$-counting of correlation functions}
Let's now consider the $N$-counting for correlation functions of local
gauge-invariant operators.
Motivated by gauge invariance and simplicity, we will consider ``single trace" operators, operators $\mathcal{O}(x)$ that look like
\begin{equation}
\mathcal{O}(x) = c(k,N) \mbox{Tr}(\Phi_1(x) ... \Phi_k(x))
\end{equation}
and which we will abbreviate as $\mbox{Tr}(\Phi^k)$.  
We will keep $k$ finite as $N \to \infty$\footnote{From the point of view of the worldsheet, these operators create closed-string excitations, such as the graviton.}.
There are two little complications here.  We must be careful about how we normalize the fields $\Phi$ and we must be careful about how we normalize the operator $\mathcal{O}$.  The normalization of the fields will continue to be such that the Lagrangian takes the form $L = \frac{1}{g^2_{YM}} \mathcal{L} = \frac{N}{\lambda} \mathcal{L}$ with $\mathcal{L}(\Phi)$ containing no explicit factors of $N$.  To fix the normalization of $\CO$ (to determine the constant $c(k,N)$) we will demand that 
when acting on the vacuum, the operator $\CO$ creates states of finite norm
in the large-$N$ limit, \ie\ 
$\langle \CO \CO \rangle_c \sim N^0$ where the subscript $c$ stands for connected.  

\begin{figure}[h]
\begin{center}
\includegraphics{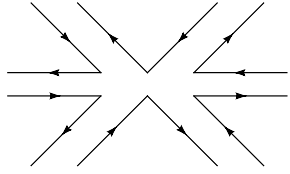}
\caption{\label{6pt}
New vertex for an operator insertion of $\mbox{Tr}(\Phi^k)$ with $k=6$}
\end{center}
\end{figure}
\begin{figure}[h]
\begin{center}
\includegraphics{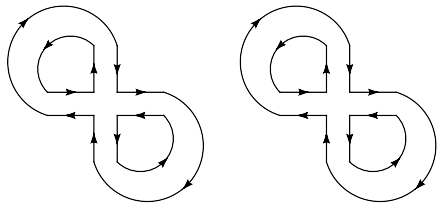}
\caption{\label{fourdisc}
Disconnected diagram contributing to the correlation function $\langle \mbox{Tr}(\Phi^4) \mbox{Tr}(\Phi^4) \rangle$}
\end{center}
\end{figure}
\begin{figure}[h]
\begin{center}
\includegraphics{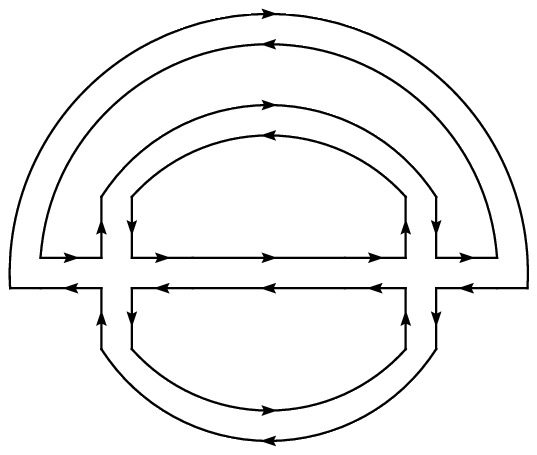}
\caption{\label{4conn}
Connected diagram contributing to the correlation function $\langle \mbox{Tr}(\Phi^4) \mbox{Tr}(\Phi^4) \rangle$}
\end{center}
\end{figure}
\begin{figure}[h]
\begin{center}
\includegraphics{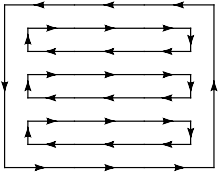}
\caption{\label{4conn_alt}
A redrawing of the connected diagram shown in Fig.~\ref{4conn}}
\end{center}
\end{figure}

To determine $c(k,N)$ we need to know how to insert single trace operators into the t'Hooft counting.  
Each single-trace operator in the correlator is a new vertex which is required to be present
in every contributing diagram.
This vertex has $k$ legs where $k$ propagators can be attached and looks like a big squid.  An example of such a new vertex appears in Fig.~\ref{6pt} which corresponds to the insertion of the operator $\mbox{Tr}(\Phi^6)$.  For the moment we don't associate any explicit factors of $N$ with the new vertex.  Let's consider the example $\langle \mbox{Tr}(\Phi^4) \mbox{Tr}(\Phi^4) \rangle$.  We need to draw two four point vertices for the two single trace operators in the correlation function.  How are we to connect these vertices with propagators?  The dominant contribution comes from disconnected diagrams like the one shown in Fig.~\ref{fourdisc}.  The leading disconnected diagram has four propagators and six index loops and so gives a factor $\lambda^4 N^2 \sim N^2$.  
On the other hand, the leading connected diagram shown in Fig.~\ref{4conn} has four propagators and four index loops and so only gives a contribution $\lambda^4 \sim N^0$.
(A way to draw the connected diagram in Fig.~\ref{4conn}
which makes the $N$-counting easier is shown in Fig.~\ref{4conn_alt} where we have deformed the two four point operator insertion vertices so that they are ``ready for contraction".)

The fact that disconnected diagrams win in the large $N$ limit is general and goes by the name ``large-$N$ factorization".  It says that single trace operators are basically classical objects in the large-$N$ limit $\langle \CO \CO \rangle \sim \langle \CO \rangle \langle \CO \rangle + O(1/N^2)$.  
\marginpar{\rotatebox{270}{\tiny [Brian Swingle]}}

The leading connected contribution to the correlation function is independent of $N$ and so $\langle \CO \CO \rangle_c \sim c^2 N^0$.  Requiring that $\langle \CO \CO \rangle_c \sim N^0$ means we can just set $c \sim N^0$.  
Having fixed the normalization of $\CO$ we can now determine the $N$-dependence of higher-order correlation functions.  For example, the leading connected diagram for $\langle \CO^3 \rangle$ where $\CO = \mbox{Tr}(\Phi^2)$ is just a triangle and contributes a factor $\lambda^3 N^{-1} \sim N^{-1}$.  In fact, quite generally we have $\langle \CO^n \rangle_c \sim N^{2-n}$ for the leading contribution.

So the operators $\CO$ (called glueballs in the 
context of QCD) create excitations of the theory that are free at large $N$ --
they interact with coupling $1/N$.
In QCD with $N=3$, quarks and gluons interact strongly, and so do their hadron composites.
The role of large-$N$ here is to make the color-neutral objects weakly interacting,
in spite of the strong interactions of the constituents.
So this is the sense in which the theory is classical:
although the dimensions of these operators can be
highly nontrivial (examples are known where they are irrational \cite{Benvenuti:2004dy}),
the dimensions of their products are additive
at leading order in $N$.

Finally, we should make a comment about the $N$-scaling of the generating
functional $Z = e^{ - W}= \vev{e^{ - N \sum_A  \lambda_A \CO^A }}$.
We've normalized the sources so that each $\lambda_A$ is an 't Hooft-like coupling,
in that it is finite as $N\to \infty$.
The effective action $W$ is the sum of connected vacuum diagrams,
which at large-$N$ is dominated by the planar diagrams.
As we've shown, their contributions go like $N^2$.
This agrees with our normalization of the gravity action,
\be S_{{\rm bulk}} \sim  { L^{d-1} \over G_N} I_{{\rm dimensionless}}
\sim N^2  I_{{\rm dimensionless}}~~~.
\ee

\subsection{Simple generalizations}
\begin{figure}[h]
\begin{center}
\includegraphics{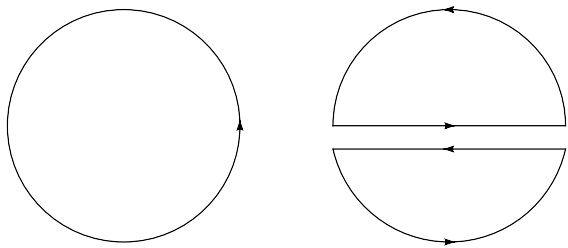}
\caption{\label{quark}
A quark vacuum bubble and a quark vacuum bubble with "gluon" exchange}
\end{center}
\end{figure}

We can generalize the analysis performed so far without too much effort.  One possibility is the addition of fields, "quarks", in the fundamental of $U(N)$.  We can add fermions $\Delta L \sim \bar{q} \gamma^{\mu} D_{\mu} q$ or bosons $\Delta L \sim |D_{\mu} q |^2$.  Because quarks are in the fundamental of $U(N)$ their propagator consists of only a single line.  When using Feynman diagrams to triangulate surfaces we now have the possibility of surfaces with boundary.  Two quark diagrams are shown in Fig.~\ref{quark} both of which triangulate a disk.  Notice in particular the presence of only a single outer line representing the quark propagator.  We can conclude that adding quarks into our theory corresponds to admitting open strings into the string theory.  We can also consider "meson" operators like $\bar{q} q$ or $\bar{q} \Phi^k q $ in addition to single trace operators.
The extension of the holographic correspondence
to include this case \cite{probebranes}
has had many applications \cite{probebraneapplications},
which are not discussed here for lack of time.
\marginpar{\rotatebox{270}{\tiny [Brian Swingle]}}

\begin{figure}[h]
\begin{center}
\includegraphics{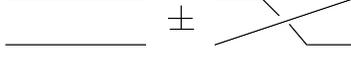}
\caption{\label{props}
Propagator for SO($N$) ($+$) or Sp($N$) ($-$) matrix models}
\end{center}
\end{figure}

Another direction for generalization is to consider different matrix groups such as $SO(N)$ or $Sp(N)$.  The adjoint of $U(N)$ is just the fundamental times the anti-fundamental.  However, the adjoint representations of $SO(N)$ and $Sp(N)$ are more complicated.  For $SO(N)$ the adjoint is given by the anti-symmetric product of two fundamentals (vectors), and for $Sp(N)$ the adjoint is the symmetric product of two fundamentals.  In both of these cases, the lines in the double-line formalism no longer have arrows.  As a consequence, the lines in the propagator for the matrix field can join directly or cross and then join as shown in Fig.~\ref{props}.  In the string language the worldsheet can now be unoriented, an example being given by a matrix field vacuum bubble where the lines cross giving rise to the worldsheet $\mathbb{RP}^2$.

\section{Vacuum CFT correlators from fields in $AdS$}
Our next goal is to evaluate $\langle e^{-\int \phi_0 \,\mathcal{O}}\rangle_{CFT} \equiv e^{-W_{CFT}[\phi_0]}$, 
where $\phi_0$ is some small perturbation around some reference value
associated with a CFT.
You may not be interested in such a quantity in itself,
but we will calculate it in a way which extends directly to 
more physically relevant quanitities (such as real-time thermal response functions).
The general form of the AdS/CFT conjecture 
for the generating funcitonal is the GKPW equation \cite{Gubser:1998bc, Witten:1998qj}
\be\langle e^{-\int \phi_0 \,\mathcal{O}}\rangle_{CFT} = Z_{{\rm strings\,in\,AdS}}[\phi_0]~.\ee This thing on the RHS is not yet a computationally effective object;
the currently-practical version of the GKPW formula is the classical limit:
\begin{equation}
\label{magic}
W_{CFT}[\phi_0]=-\ln\langle e^{\int \phi_0 \mathcal{O}}\rangle_{CFT} \simeq {\rm extremum}_{``\phi|_{z=\epsilon}\sim \phi_0"}\left(\hl{N^2} I_{grav}\left[\phi\right]\right) + O\left(\frac{1}{\hl{N^2}}\right) + O\left(\frac{1}{\sqrt{\lambda}}\right)
\end{equation}

There are many things to say about this formula.
\begin{itemize}
\item In the case of matrix theories like those described in the previous section,
the classical gravity description is valid for large $\hl{N}$ and large $\lambda$.  
In some examples there is only one parameter which controls the validity of the gravity 
description.
In (\ref{magic}) we've made the $\hl{N}$-dependence
explicit: in units of the $AdS$ radius, the Newton constant is $ {L^{d-1}\over G_N} = \hl{N^2} $ . $I_{grav}$ is some dimensionless
action.

%
\item We said that we are going to think of $\phi_0$ as a small perturbation.
Let us then make a perturbative expansion in powers of $\phi_0$:
\begin{equation}
W_{CFT}[\phi_0]=W_{CFT}[0]+\int d^D x ~~
\phi_0 (x) G_1(x) + \frac{1}{2} \int \int d^D x_1 d^D x_2 ~ \phi_0 (x_1) \phi_0 (x_2) G_2 (x_1,x_2) +...
\end{equation}
where
\begin{eqnarray}
G_1(x) = \langle \CO(x) \rangle= \frac{\delta W}{\delta \phi_0(x)}\vert_{\phi_0 = 0}, \\
G_2 (x) = \langle \CO(x_1) \CO(x_2) \rangle_c = \frac{\delta^2 W}{\delta \phi_0 (x_1) \delta \phi_0 (x_2)} \vert_{\phi_0=0}.
\end{eqnarray}
Now if there is no instability, then $\phi_0$ is small implies $\phi$ is small.
For one thing, this means that we can 
ignore interactions of the bulk fields in computing two-point functions.
For $n$-point functions, we will need to know terms in the bulk action of 
degree up to $n$ in the fields. 

\item
Anticipating divergences at $z\rightarrow 0$, we have introduced a cutoff 
in (\ref{magic}) (which will be a UV cutoff in the CFT) and 
set boundary conditions at $z=\epsilon$.  
They are in quotes because they require a bit of refinement (this will happen in subsection \ref{sec:refinement}).

\item
Eqn (\ref{magic}) is written as if there is just one field in the bulk.  Really there is a $\phi$ for every operator $\mathcal{O}$
in the dual field theory. For such a pair, we'll say `$\phi$ couples to $\mathcal{O}$' at the boundary. How to match up
fields in the bulk and operators in the QFT? 
In general this is hard and information from string theory is useful. 
Without specifying a definite field theory, we can say a few general things:
\begin{enumerate}
\item We can organize the both hand side into representations of the conformal group.
In fact only conformal primary operators correspond to `elementary fields' in the 
gravity action, and their descendants correspond to derivatives of those fields.
More about this loaded word `elementary' in a moment.

\item Only `single-trace' operators (like the $\tr \Phi^k$s of the previous section) 
correspond to `elementary fields' $\phi$ in the bulk.
The excitations created by multi-trace operators (like $\(\tr \Phi^k\)^2$)
correspond to multi-particle states of $\phi$ (in this example, a 2-particle state).
Here I should stop and emphasize that this word `elementary' is well-defined because
we have assumed that we have a weakly-coupled theory in the bulk,
and hence the Hilbert space is approximately a Fock space, organized
according to the number of particles in the bulk.
A well-defined notion of single-particle state depends on large-$\hl{N}$ -- 
if $\hl{N}$ is not large, it's not true that 
the overlap between $\tr \Phi^2 \tr \Phi^2\ket{0} $ and 
$\tr \Phi^4 \ket{0}$ is small\footnote{It is clear that the 't Hooft limit
is not the only way to achieve such a situation, but I am using the 
language specific to it because it is the one I understand.}.

\item There are some simple examples of the correspondence between
bulk fields and boundary operators that are determined by symmetry.
The stress-energy tensor $T_{\mu\nu}$ is the response of a local QFT to local change in the metric, $S_{bdy} \ni \int \gamma_{\mu\nu}T^{\mu\nu}$. 
Here we are writing $\gamma_{\mu\nu}$ for the metric on the boundary.
In this case
\begin{equation}
g_{\mu\nu} \leftrightarrow T_{\mu\nu}~~.
\end{equation}
Gauge fields in the bulk correspond to 
currents in the boundary theory:
\begin{equation}
A^{a}_{\mu} \leftrightarrow J^{\mu}_{a}
\end{equation}
\ie\ $S_{bdy}\ni \int A^{a}_{\mu}J^{\mu}_{a}$.
We say this mostly because we can contract all the indices 
to make a singlet action.
In the special case where the gauge field is massless, the current is conserved.
\end{enumerate}

\item Finally, something that needs to be emphasized is that 
changing the {\it Lagrangian} of the CFT
(by changing $\phi_0$) is accomplished
by changing the {\it boundary condition} in the bulk.  
The bulk equations of motion 
remain the same (\eg\ the masses of the bulk fields don't change).
This means that actually changing the bulk action
corresponds to something more drastic in the boundary theory.  
One context in which
it is useful to think about varying the bulk coupling 
constant is in thinking about the renormalization group.
We motivated the form $\int\(  2 \Lambda + \CR + ....\)$ 
of the bulk action by Wilsonian naturalness, 
which is usually enforced by the RG,
so this is a delicate point.
For example, soon we will compute 
the ratio of the shear viscosity to the entropy density, $\eta \over s$,
for the plasma made from {\it any CFT} that has an Einstein gravity dual;
the answer is always $1 \over 4\pi$.
Each such CFT is what we usually think of as a universality class,
since it will have some basin of attraction in the space of nearby QFT couplings.
Here we are saying that a whole {\it class} of universality classes 
exhibits the same behavior.

What's special about these theories from the QFT point of view?
Our understanding of this `bulk universality' is obscured by 
our ignorance about quantum mechanics in the bulk.
Physicists with what could be called a monovacuist inclination
may say that what's special about them is that they exist\footnote{
{\bf Monovacuist} (n):
One who believes that a theory of quantum gravity
should have a unique groundstate
(in spite of the fact that we know many examples of much simpler systems
which have many groundstates, and in spite of all the 
evidence to the contrary (\eg\ \cite{Silverstein:2004id, Douglas:2006es})).
}.
The issue, however, arises for interactions in the bulk which are 
quite a bit less contentious than gravity, so this seems unlikely to me to be the answer.

\end{itemize}

\subsection{Wave equation near the boundary and dimensions of operators}
\label{sec:refinement}
\def\kkk{{\mathfrak{K}}}

The metric of AdS 
(in Poincar\'e coordinates, so that the constant-$z$ slices are just 
copies of Minkowski space) 
is
\begin{equation}
ds^2=L^2\frac{dz^2+dx^{\mu}\,dx_{\mu}}{z^2}\equiv g_{AB}dx^{A}dx^{B} ~~~~~~~
A=0, \ldots , d, ~~~x^A=(z,x^\mu)~~.
\end{equation}
As the simplest case to consider, let's think about a scalar field in the bulk.
An action for such a scalar field suggested by Naturalness is
\begin{equation}\label{scalaraction}
S=-\frac{\kkk}{2} \int d^{d+1}x \sqrt{g} \left[
g^{AB}\partial_{A}\phi\partial_{B}\phi + 
m^2\phi^2 + b \phi^3 + \ldots 
\right]~.
\end{equation}
Here $\kkk$ is just a normalization constant; we are assuming that 
the theory of $\phi$ is weakly coupled and one may think of $\kkk$
as proportional to $\hl{N^2}$.
For this metric $\sqrt{g}=\sqrt{|\det\,g|}=\left(\frac{L}{z}\right)^{d+1}$.  
Our immediate goal is to compute a two-point function of the operator $\CO$
to which $\phi$ couples, so we will ignore the interaction terms in 
(\ref{scalaraction}) for a while.
Since $\phi$ is a scalar field we can rewrite the kinetic term as 
\begin{equation}
g^{AB}\partial_{A}\phi\partial_{B}\phi=\left(\partial \phi\right)^2= g^{AB}D_{A}\phi D_{B}\phi
\end{equation}
where $D_{A}$ is the covariant derivative,
which has the nice property that $D_{A}(g_{BC})=0$, so we can
move the $D$s around the $g$s with impunity. By integrating by parts we can rewrite the action in a useful way:
\begin{equation}
S=-\frac{\kkk}{2}\int d^{d+1}x \left[\partial_{A}\left(\sqrt{g}\,g^{AB}\phi\,\partial_{B}\phi\right)-\phi\,\partial_{A}\left(\sqrt{g}\,g^{AB}\,\partial_{B}\phi\right)+\sqrt{g}\left(m^2\phi^2+\ldots\right)\right]
\end{equation}
and finally by using Stokes' theorem we can rewrite the action as
\begin{equation}\label{IBP}
S=-\frac{\kkk}{2}\int_{\partial AdS} d^{d}x\,\sqrt{g}\,g^{zB}\phi\,\partial_{B}\phi - \frac{\kkk}{2}\int \sqrt{g}\,\phi\left(-\Box+m^2\right)\phi+ O(\phi^3)
\end{equation}
where we define the scalar 
Laplacian $\Box\phi=\frac{1}{\sqrt{g}}\partial_{A}\left(\sqrt{g}\,g^{AB}\,\partial_{B}\right)\phi=D^{A}D_{A}\phi$.
Note that we wrote all these covariant expressions without ever introducing
Christoffel symbols.

We can rewrite the boundary term more covariantly as
\begin{equation}
\int_{\mathcal{M}}\sqrt{g}\,D_{A}J^{A} = \int_{\partial \mathcal{M}}\sqrt{\gamma}\,n_{A}J^{A}~.
\end{equation}
The metric tensor $\gamma$ is defined as
\begin{equation}
ds^2|_{z=\epsilon} \equiv \gamma_{\mu\nu}dx^{\mu}dx^{\nu} = {L^2 \over \epsilon^2} \eta_{\mu\nu}dx^\mu dx^\nu
\end{equation}
\ie\ it is the induced metric on the boundary surface $z=\epsilon$.
The vector $n_{A}$ is a unit vector normal to boundary ($z=\epsilon$). We can find an explicit expression for it
\begin{equation}
n_{A}\propto \frac{\partial}{\partial z} ~~~~ g_{AB}n^{A}n^{B}|_{z=\epsilon}=1 ~~~~~~~~~~ \Rightarrow  ~~~~~~~~~~ n=\frac{1}{\sqrt{g_{zz}}}\frac{\partial}{\partial z}=\frac{z}{L}\frac{\partial}{\partial z}~~.
\end{equation}

From this discussion we have learned the following:
\begin{itemize}
\item the equation of motion for small fluctuations of $\phi$ 
is $ (- \Box + m^2) \underline \phi = 0 $~~\footnote{We will use an underline to denote
fields which solve the equations of motion.}.
\item If $\underline \phi$ solves the equation of motion,
the on-shell action is just given by the boundary term.
\end{itemize} 

Next we will derive the promised formula
relating bulk masses and operator dimensions 
$$\Delta (\Delta - d) = m^2 L^2 $$
by studying the 
AdS wave equation near the boundary.

Let's take advantage of translational invariance in $d$ dimensions, $x^\mu \rightarrow x^\mu + a^\mu$, to Fourier decompose the scalar field:
\begin{equation} \label{eq2}
\phi (z, x^\mu) = e^{i k_\mu x^\mu} f_k (z), ~~~~ k_\mu x^\mu \equiv - \omega t + \vec k \cdot \vec x~.
\end{equation} 
In the Fourier basis, substituting (\ref{eq2}) into the wave equation
$(- \Box +m^2) \phi = 0 $ and 
using the fact that the metric only depends on $z$, the wave equation is:
\begin{eqnarray} 
0&=& (g^{\mu \nu} k_{\mu} k_{\nu} -\frac{1}{\sqrt{g}} \partial_z (\sqrt{g} g^{zz} \partial_z) + m^2) f_k (z) \\
\label{eq3}
&= &\frac{1}{L^2} [ z^2 k^2 - z^{d+1} \partial_z (z^{-d+1} \partial_z) +m^2 L^2] f_k(z) ,
\end{eqnarray}
where we have used $g^{\mu \nu} = (z/L)^2 \delta^{\mu \nu}$. The solutions of (\ref{eq3}) are Bessel functions; we can learn a lot without using that information. 
For example, look at the solutions near the boundary 
(\ie\  $z \rightarrow 0$). In this limit we have power law solutions, which are spoiled by the $z^2 k^2$ term. To see this, try using $f_k = z^\Delta$ in  (\ref{eq3}):
\begin{eqnarray}
0&=&k^2 z^{2+\Delta} - z^{d+1} \partial_z(\Delta z^{-d+\Delta}) +m^2 L^2 z^\Delta \\
&=&  (k^2 z^2 - \Delta (\Delta -d) +m^2 L^2) z^\Delta,
\end{eqnarray}
and for $z \rightarrow 0$ we get:
\begin{equation} \label{eq4}
\Delta (\Delta -d) = m^2 L^2~~.
\end{equation}
The two roots of (\ref{eq4}) are
\begin{equation}
\Delta_\pm = \frac{d}{2} \pm \sqrt{\left( \frac{d}{2}\right)^2 +m^2 L^2}.
\end{equation}

\underline{Comments}
\begin{itemize}
\item The solution proportional to $z^{\Delta_-}$ is bigger near $z \rightarrow 0$.
\item $\Delta_+ > 0~~\forall~~m$, therefore $z^{\Delta_+}$ decays near the boundary
for any value of the mass.
\item $\Delta_+ + \Delta_- = d$.
\end{itemize}
We want to impose boundary conditions that allow solutions. 
Since the leading behavior near the boundary of a generic solution
is $\phi \sim z^{\Delta_-}$, we impose
\begin{equation}
\phi (x,z) |_{z= \epsilon} = \phi_0 (x,\epsilon) = \epsilon^{\Delta_-} \phi_0 ^{Ren}(x),
\end{equation}
where $\phi_0^{Ren}$ is a renormalized source field. 
With this boundary condition $\phi_0^{Ren}$ is a finite quantity in the limit $\epsilon \rightarrow 0$.

\underline{Wavefunction renormalization of $\CO$} (Heuristic but useful)

Suppose:
\begin{eqnarray}
S_{bdy}&\ni& \int_{z=\epsilon} d^{d} x ~ \sqrt{\gamma_{\epsilon}} ~\phi_0 (x,\epsilon) \CO(x,\epsilon) \\
&= &\int d^d x~ \left(\frac{L}{\epsilon}\right)^d (\epsilon^{\Delta_-} \phi_0^{Ren}(x)) \CO(x,\epsilon),
\end{eqnarray}
where we have used $\sqrt{\gamma} = (L/\epsilon)^d$. Demanding this to be finite as $\epsilon \rightarrow 0$ we get:
\begin{eqnarray}
\CO(x,\epsilon) &\sim& \epsilon^{d-\Delta_-} \CO^{Ren}(x) \\
&=&\epsilon^{\Delta_+} \CO^{Ren}(x),
\end{eqnarray}
where in the last line we have used $\Delta_+ + \Delta_- = d$. Therefore, the scaling dimension of $\CO^{Ren}$ is $\Delta_+ \equiv \Delta$. 
We will soon see that $\langle \CO(x)\CO(0) \rangle \sim \frac{1}{|x|^{2 \Delta}}$,
confirming that $\Delta$ is indeed the scaling dimension.


We are solving a second order ODE, therefore we need two conditions in order to determine a solution (for each $k$). So far we have imposed one condition at the 
boundary of AdS:\\
$~~~\bullet$ For $z \rightarrow \epsilon,~~\phi \sim z^{\Delta_-} \phi_0~ +$ (terms subleading in $z$).\\
In the Euclidean case (we discuss real time in the next subsection), 
we will also impose\\
$~~~\bullet$ $\phi$ regular in the interior of AdS (\ie\  at $z \rightarrow \infty$).

\underline{Comments on $\Delta$}

\begin{enumerate}
\item The $\epsilon^{\Delta_-}$ factor is independent of $k$ and $x$, which is a consequence of a local QFT (this fails in exotic examples).
\item \underline{Relevantness}:
If $m^2 > 0$: This implies $\Delta \equiv \Delta_+ >d$, so $\CO_\Delta$ is an irrelevant operator. This means that if you perturb the CFT by adding $\CO_\Delta$ to the Lagrangian, then its coefficient is some mass scale to a negative power:
\begin{equation}
\Delta S = \int d^d x ~(\mathrm{mass})^{d-\Delta} \CO_\Delta,
\end{equation}
where the exponent is negative, so the effects of such an operator go away in the IR, 
at energies $E < {\rm mass}$. 
$\phi \sim z^{\Delta_-} \phi_0$ {\it is} this coupling.  It grows in the UV (small $z$).
If $\phi_0$ is a finite perturbation, it will back-react on the metric and 
destroy the asymptotic AdS-ness of the geometry: extra data about the UV will be required.

$m^2 = 0 \leftrightarrow \Delta = d$ means that $\CO$ is marginal.  

If $m^2<0$, then $\Delta<d$, so $\CO$ is a relevant operator.
Note that in $AdS$, $m^2 < 0$ is ok if $m^2$ is not too negative.
Such fields 
with $m^2 > -|m_{BF}|^2 \equiv - (d/2 L)^2$ are called 
``Breitenlohner-Freedman (BF)-allowed tachyons".  
The reason you might think that $m^2 < 0$ is bad is that usually it means
an instability of the vacuum at $\phi=0$.  
An instability occurs when a normalizable mode grows with time without a source. 
But for $m^2 <0$, $\phi \sim z^{\Delta_-}$ decays near the boundary (\ie\  in the UV). 
%
This requires a gradient energy of order $\sim\frac{1}{L}$, which 
can stop the field from condensing.

To see what's too negative, 
consider the formula for the dimension, 
$\Delta_\pm = {d\over 2}
\pm \sqrt{ \( d \over 2\)^2 + m^2 L^2 }.   $
For $m^2 < m_{BF}^2$, the dimension becomes complex.

\item The formula relating the mass of a bulk field
and the dimension of the associated operator depends on their spin. 
For example, for a massive gauge field in $AdS$ with action
$$S = - \int_{AdS}\( {1\over 4}  F_{\mu\nu}F^{\mu\nu} - \half m^2 A_\mu A^\mu \), $$
the boundary behavior of the wave equation implies that  $A_\mu \buildrel{z\to0}\over{\sim} 
z^\alpha$ with
\begin{equation}
\alpha (\alpha-d+2)=m^2L^2.
\end{equation}
For the particular case of $A_\mu$ massless this can be seen to lead to
$
\Delta(j^\mu) = d-1,
$
which is the dimension of a conserved current in a CFT.
Also, the fact that $g_{\mu \nu}$ is massless implies:
\begin{equation}
\Delta(T^{\mu \nu}) = d,
\end{equation}
which is required by conformal Ward identities.

\end{enumerate}

\subsection{Solutions of the $AdS$ wave equation and real-time issues}

An approach which uses the symmetries of $AdS$ \cite{Witten:1998qj}
is appealing.
However, it 
isn't always available (for example if there is a black hole in the spacetime). 
Also, it can be misleading:
as in quantum field theory, 
scale-invariance is inevitably broken by the UV cutoff.





Return to the scalar wave equation in momentum space:
\be
\label{star}
0 = [z^{d+1} \p_z (z^{-d+1}\p_z) -m^2 L^2 - z^2 k^2] f_k (z) ~~.
\ee
We will now stop being agnostic about the signature and 
confront some issues that arise for real time correlators.
If $k^2 >0$, that is, $k^\mu$ is spacelike (or Euclidean), 
the solution is
\be
 f_k(z) = A_K z^{d/2} K_{\nu} (kz) + A_I z^{d/2} I_{\nu} (kz)
\ee
where $\nu = \Delta - d/2 = \sqrt{(d/2)^2 +m^2 L^2}$.
In the interior of $AdS$ ($z \to \infty$), the Bessel functions behave as 
\be
 K_\nu (kz) \sim e^{-kz} \qquad I_\nu (kz)  \sim e^{kz} ~.
\ee
So we see that the regularity in the interior uniquely fixes $\underline f_k$
and hence the bulk-to-boundary propagator.
Actually there is a theorem (the Graham-Lee theorem) addressing this
issue for gravity fields (and not just linearly in the fluctuations);
it states that if you specify a Euclidean metric on the boundary
of a Euclidean AdS (which we can think of as topologically a $S^d$ by 
adding the point at infinity in $\IR^d$)
modulo conformal rescaling, then the metric on the space inside
of the $S^d$ 
is uniquely determined.
A similar result holds for gauge fields.

In contrast to this, in Lorentzian signature with timelike $k^2$, \ie\  for on-shell states with
$\omega^2 > \vec{k}^2$, there exist 
two linearly independent solutions 
with the same leading behavior at the UV boundary.
In terms of $q\equiv \sqrt{\omega^2 - \vec{k}^2}$, the solutions are
\be
 z^{d/2} K_{\pm \nu} (iqz) \sim  e^{\pm iqz} \quad (z \rightarrow \infty)
\ee
so these modes oscillate 
in the far IR region of the geometry\footnote
{This $z\to \infty$, far IR region of the geometry is called the `Poincar\'e horizon'.
A few words of clarification may be useful regarding this terminology.  
The form \eqref{AdSmetric} of the AdS metric that we have been discussing
is not geodesically complete.  If we follow all the geodesics to their 
logical conclusions, the geometry we find 
is called `global AdS'; it has constant-$z$ spatial sections which are $d-1$-spheres.
The 
coordinates we have been using 
in \eqref{AdSmetric} cover a subset of this geometry called the `Poincar\'e patch'.
The Poincar\'e horizon at $z \to \infty$ represents a boundary of this coordinate patch.
}.
This ambiguity reflects the many possibilities for real-time Green's functions in the QFT.
One useful choice is the retarded Green's function, which describes 
causal response of the system to a perturbation.
This choice corresponds to a choice of boundary conditions 
at $z \to \infty$ describing stuff falling into the horizon \cite{Son:2002sd},
\ie\ moving towards larger $z$ as time passes.
There are three kinds of reasons for this prescription\footnote{This is not yet
the complete prescription for computing retarded functions; the other ingredient
will appear in subsection \ref{sec:response}.}: 
\begin{itemize}
\item Both the retarded Green's functions and stuff falling through the horizon
describe things that {\it happen}, rather than {\it unhappen}.

\item You can check that this prescription gives the correct analytic
structure of $G_R(\omega)$ (\cite{Son:2002sd} and all the hundreds
of papers that have used this prescription).

\item It has been derived from a holographic version of the Schwinger-Keldysh 
prescription \cite{Herzog:2002pc, Maldacena:2001kr, Skenderis:2008dh}.
\end{itemize}
The fact that stuff goes past the horizon and doesn't come out 
is what breaks time-reversal invariance in the holographic computation
of $G^R$.  Here, the ingoing choice is 
$\phi(t, z) \sim e^{ - i \omega t + i q z} $,
since as $t$ grows, the wavefront moves to larger $z$.
This specifies the solution which computes the causal response
to be $z^{d/2} K_{+\nu}(iqz)$.

The same prescription, adapted to the black hole horizon, will work in the finite temperature case.
%

One last thing we must deal with before proceeding is
to define what we mean by a `normalizable' mode, or solution,
when we say that we have many normalizable solutions
for $k^2<0$ with a given scaling behavior.
In Euclidean space, $\phi$ is normalizable when $S[\phi]<\infty$.
This is because when we are thinking about the partition function $Z[\phi]=\sum_{\phi} e^{-S[\phi]}$,
modes with boundary conditions which force $S[\phi] = \infty$
would not contribute.
In real-time, we say $\phi$ is normalizable if $E[\phi]<\infty$ where,
\be
 E[\phi] = \int_{\Sigma} d^{d-1} x dz \sqrt{h} T_{\mu \nu} [\phi] n^\mu \xi^\nu
 = \int_{x^0 =\text{constant}} d^{d-1} x dz \sqrt{h}  T^0_{~0} [\phi]
\ee
where $\Sigma$ is a given spatial slice, $h$ is the induced metric on that slice,
$n^\mu$ is a normal unit vector to $\Sigma$ and $\xi^\mu$ is a timelike killing vector.
$T_{AB}$ is defined as
\be
 T_{AB} [\phi] = -{2 \over \sqrt{g}} {\delta \over \delta g^{AB}} S_{Bulk} [\phi] .
\ee

\subsection{Bulk-to-boundary propagator in momentum space}
\label{sec:bulktoboundary}

We return to considering spacelike $k$ in this section.
Let's normalize our solution of the wave equation 
by the condition $f_k (z=\epsilon) =1$.
This means that its Fourier transform is a $\delta$-function in position space 
$\delta^d(x)$
when evaluated at
$z=\epsilon$, not at the actual boundary $z=0$.
The solution, which we can call the `regulated bulk-to-boundary propagator', is then 
\be
 \underline f_k(z) = {z^{d/2} K_\nu(kz) \over \epsilon^{d/2} K_\nu (k \epsilon)}.
\ee
\def\ub{\underline}
The general position space solution can be obtained by Fourier decomposition
\be\label{bulktoboundary}
 \ub{\phi}^{[\phi_0]} (x) = \int d^d k e^{ikx} \ub{f}_k (z) \phi_0 (k,\epsilon)~.
\ee
The `on-shell action' (\ie\ the action evaluated on the saddle-point solution) is
(using (\ref{IBP}))
\begin{align*}
\begin{split}
 S[\ub{\phi}] &= -{\kkk \over 2} \int d^d x \sqrt{\gamma} \ub{\phi} n \cdot \p \ub{\phi} \\
              &= -{\kkk \over 2} \int d^d x \int d^d k_1 \int d^d k_2
                 e^{i (k_1 + k_2)x} \phi_0 (k_1,\epsilon) \phi_0 (k_2,\epsilon) L^{d-1} z^{-d} \ub f_{k_1} (z) z \p_z \ub f_{k_2} (z) |_{z=\epsilon}\\
              &= -{\kkk L^{d-1} \over 2} \int d^d k \phi_0 (k,\epsilon) \phi_0 (-k,\epsilon) \CF_\epsilon (k)
\end{split}
\end{align*}
and therefore\footnote{If you don't like functional derivatives, you can see
(\ref{functionalderiv}) by calculating
\be
 \vev{\CO (k_1 ) \CO (k_2)}^\epsilon_c
 = \( {\p^2 \over \p \lambda_1 \p \lambda_2} W[\phi_0(x) = \lambda_1 e^{ik_1 x} + \lambda_2 e^{ik_2 x}] \) |_{\lambda_1=\lambda_2=0}~.
\ee
},
\be\label{functionalderiv}
 \vev{\CO (k_1 ) \CO (k_2)}^\epsilon_c
 = -{\delta \over \delta \phi_0 (k_1)}{\delta \over \delta \phi_0 (k_2)} S
 = (2\pi)^d \delta^d(k_1+k_2)  \CF_\epsilon (k_1)~.
\ee
Here $\CF_\epsilon (k)$ (sometimes called the `flux factor') is 
\be\label{fluxfactor}
 \CF_\epsilon (k) =  z^{-d} \ub f_{-k} (z) z \p_z \ub f_k (z) |_{z=\epsilon} + (k \leftrightarrow -k)
 = 2 \epsilon^{-d+1} \p_z \( {z^{d/2} K_\nu (kz) \over \epsilon^{d/2} K_\nu (\epsilon z)} \) |_{z=\epsilon}~.
\ee
The small-$u$ (near boundary) behavior of $K_\nu(u)$ is
\begin{align*}
\begin{split}
K_\nu (u) = &u^{-\nu} (a_0 + a_1 u^2 + a_2 u^4 + \cdots ) ~~~~~~~(leading~term)\\
           +&u^{\nu}\hl{\ln u} (b_0 + b_1 u^2 +b_2 u^4 + \cdots ) ~~~~(subleading~term)
\end{split}
\end{align*}
where the coefficients of the series $a_i,~b_i$ depend on $\nu$.
For non-integer $\nu$, there would be no $\hl{\ln u}$ in the second line,
and so we make it pink.
Of course, we saw in the previous subsection (with very little work)
that {\it any} solution of the bulk wave equation has this kind of form
(the boundary is a regular singular point of the ODE).
We could determine the $a$s and $b$s recursively by the same procedure.
This is just like a scattering problem in 1d quantum mechanics.
The point of the Bessel function here is to choose
which values of the coefficients $a, b$ give a function 
which has the correct behavior at the other end, \ie\ at $z \to \infty$.  
Plugging the asymptotic expansion of the Bessel function into (\ref{fluxfactor}),
\begin{eqnarray}
\label{bovera}
 \CF_\epsilon (k) &=& 2 \epsilon^{-d+1} \p_z
 \( { (kz)^{-\nu+d/2} (a_0 + \cdots ) +  (kz)^{\nu+d/2} \hl{\ln kz} (b_0 + \cdots )
 \over (k\epsilon)^{-\nu+d/2} (a_0 + \cdots ) +  (k\epsilon)^{\nu+d/2} \hl{\ln k\epsilon} (b_0 + \cdots )}
 \)|_{z=\epsilon} \\ \nonumber
 &=& 2\epsilon^{-d} \left[ \left\{ {d \over 2} - \nu (1+ c_2 (\epsilon^2 k^2) + c_4 (\epsilon^4 k^4) + \cdots) \right\}
 + \left\{ \nu {2 b_0 \over a_0}  (\epsilon k)^{2 \nu} \hl{\ln (\epsilon k)} (1 + d_2 (\epsilon k)^2 + \cdots) \right\} \right] \\ \nonumber
 &\equiv&{\rm (I)} + {\rm (II)}
\end{eqnarray}
where (I) and (II) denote the first and second group of terms of the previous line.

(I) is a Laurent series in $\epsilon$ with coefficients
which are positive powers of $k$ (\ie\  analytic in $k$ at $k=0$).
These are contact terms, \ie\  short distance goo that we don't care about
and can subtract off.
We can see this by noting that 
\be
 \int d^d k e^{-ikx} (\epsilon k)^{2m} \epsilon^{-d} = \epsilon^{2m-d} \Box^m_x \delta^d (x) 
\ee
for $m>0$. The $\epsilon^{2m-d}$ factor reinforces the notion that
$\epsilon$, which is an IR cutoff in AdS, is a UV cutoff for the QFT.

The interesting bit of $\CF (k)$, which gives the $x_1 \neq x_2$ behavior of the correlator 
(\ref{functionalderiv}), is
non-analytic in $k$:
\be
 {\rm (II)} = - 
 2\nu \cdot {b_0 \over a_0} k^{2\nu} \hl{\ln (k\epsilon)} \cdot \epsilon^{2\nu-d} (1 + \CO (\epsilon^2)),
 \qquad \left( {b_0 \over a_0} = {(-1)^{\nu-1} \over 2^{2\nu} \nu  \Gamma(\nu)^2} ~\text{ for }~\nu \in \mathbb{Z} \right).
\ee
To get the factor of $2\nu$, one must expand
both the numerator and the denominator in (\ref{bovera});
this important subtlety was pointed out in \cite{D'Hoker:1999pj}\footnote{
The correctness of 
this normalization of the two-point function can be verified by computing 
a {\it three} point function $\vev{J_\mu \CO^\dagger \CO}$ (where $J$ is a 
conserved current under which $\CO$ is charged) and using the
Ward identity for $J$-conservation to relate this to $\vev{\CO^\dagger \CO}$ \cite{D'Hoker:1999pj}.
}.
The Fourier transformation of the leading term of (II) is given by
\be
 \int d^d k e^{-ikx} {\rm (II)} = 
 {2 \nu \Gamma (\Delta_+) \over \pi^{d/2} \Gamma(\Delta_+-d/2)} {1 \over x^{2\Delta_+}} \epsilon^{2\nu-d}~.
\ee
Note that the $AdS$ radius appears only through the overall normalization of the 
correlator (\ref{functionalderiv}), in the combination $\kkk L^{d-1}$.

%
Now let's deal with the pesky cutoff dependence.  Since $\epsilon^{2\nu-d} = \epsilon^{- 2\Delta_-}$ if we let
$\phi_0 (k, \epsilon) = \phi_0^{\text{Ren}}(k)\epsilon^{\Delta_-}$ as before,
the operation
\be
 {\delta \over \delta \phi_0 (k, \epsilon)} = \epsilon^{-\Delta_-} {\delta \over \delta \phi_0^{\text{Ren}} (k)}
\ee
removes the potentially-divergent factor of $\epsilon^{-2\Delta_-}$. We also see that for $\epsilon \rightarrow 0$, the $\CO(\epsilon^2)$
terms vanish.
%


If you are bothered by the infinite contact terms (I), 
there is a prescription to cancel them, appropriately 
called Holographic Renormalization \cite{Skenderis}. 
Add to $S_{bulk}$ the local, intrinsic boundary term 
\begin{eqnarray*}
\Delta S = S_{\mbox{\tiny{c.t.}}}& =& \frac{\kkk}{2}\int_{\mbox{\tiny{bdy}}}d^dx\left(-\Delta_-
L^{d-1}\epsilon^{2\Delta_- - d}\left(\phi_0^{\mbox{\tiny{Ren}}}(x)\right)^2\right)\\
& =&-\Delta_- \frac{\kkk}{2L}\int_{\partial AdS, z=\epsilon}\sqrt{\gamma}\,\phi^2(z,x)
\end{eqnarray*}
and redo the preceding calculation.
Note that this doesn't affect the equations of motion, nor does it affect $G_2(x_1\neq x_2)$.

\subsection{The response of the system
to an arbitrary source}
\label{sec:response}

Next we will derive a very important formula for the response of the system
to an arbitrary source.  The preceding business with the on-shell action is cumbersome,
and is inconvenient for computing real-time Green's functions.
The following result \cite{Balasubramanian:1998de, Klebanov:1999tb, 
Papadimitriou:2004ap, Iqbal:2008by}
circumvents this confusion.

The solution of the equations of motion, satisfying the conditions we want 
in the interior of the geometry, behaves near the boundary as
\be 
\label{solutionatboundary}
\phi(z, x) \approx 
\( {z\over L} \) ^{\Delta_-} \phi_0 (x) \( 1 + \CO(z^2)\) +
\( {z\over L} \) ^{\Delta_+} \phi_1 (x) \( 1 + \CO(z^2)\)  ; \ee
this formula defines the coefficient $\phi_1$ of the subleading behavior
of the solution.
First we recall some facts from classical mechanics.
Consider the dynamics of a particle in one dimension, with action
$S[x] = \int_{t_i}^{t_f} dt L(x(t), \dot x(t))$.
The variation of the action with respect to the initial value of the coordinate 
is the momentum:
\be\label{HJeqn}
 {\delta S \over \delta x(t_i) } = \Pi(t_i) , ~~~ \Pi(t)  \equiv  {\partial L \over \partial \dot x}~~. \ee
Thinking of the radial direction of $AdS$ as time, the following is a mild generalization of 
(\ref{HJeqn}):
\be 
\label{responseformula}
\vev{\CO(x)}
= {\delta W [\phi_0]\over \delta \phi_0(x)} 
= \lim_{z\to0} \( z\over L\)^{\Delta_-} \Pi(z,x) |_{{\rm finite}},\ee
where $\Pi \equiv {\partial \CL \over \partial (\partial_z \phi) } $
is the bulk field-momentum, with $z$ thought of as time.
There are two minor subtleties: \\
(1) the factor of $z^\Delta_-$ arises because $\phi_0$ differs 
from the boundary value of $\phi$ by a factor: $\phi \sim z^{\Delta_-} \phi_0$,
so ${\partial \over \partial \phi_0} = z^{-\Delta_-}{ \partial \over \partial \phi(z=\epsilon)}$.\\
(2) $\Pi$ itself in general (for $m\neq 0$) has a term proportional to the source $\phi_0$,
which diverges near the boundary; this is related to the contact terms
in the previous description.  Don't include these terms in (\ref{responseformula}).
Finally, then, we can evaluate the field momentum 
in the solution (\ref{solutionatboundary}) and find\footnote{
This formula is not correct on the support of the source 
$\phi_0$.  
If one wants to use this formula to 
compute expectation values 
with a nonzero source (and not just correlation functions at finite separation),
the terms proportional to the source must be included and renormalized.
For cautionary examples and their correct treatment, please see \cite{Bianchi:2001de}.  
Thanks to Kostas Skenderis for emphasizing this point.
}
\be \label{response} \vev{\CO(x)} = \kkk { 2 \Delta - d \over L} \phi_1(x). \ee
This is an extremely important formula.
We already knew that the leading behavior of the solution
encoded the source in the dual theory, 
\ie\ the perturbation of the {\it action} of the QFT.
Now we see that the coefficient of the subleading falloff 
encodes the response \cite{Balasubramanian:1998de}.  
It tells us how the 
{\it state} of the QFT changes as a result of the perturbation\footnote{
An important practical remark: In general, the bulk system will be nonlinear -- a 
finite perturbation of $\phi_0$ will source 
other bulk fields, such as the metric.  Actually finding the 
resulting $\phi(z,x)$ in (\ref{solutionatboundary}) may be quite complicated.
The prescription for extracting the response, given that solution,
is always as described here.
}.

This formula applies in the real-time case \cite{Iqbal:2009fd}.  
For example, to describe linear response, 
$\delta \vev{\CO} = \delta \phi_0 G + O(\delta \phi_0)^2) $,
then (\ref{response}) says that 
\be G(\omega, k) = \kkk { 2 \Delta - d \over L }  { \phi_1(\omega ,k) \over \phi_0(\omega,k)}. \ee
Which kind of Green's function we get 
depends on the boundary conditions we impose in the interior.

\subsection{A useful visualization}

We are doing classical field theory in the bulk, \ie\ solving a boundary value problem. 
We can describe the 
expansion about an extremum of classical action 
in powers of $\phi_0$
in terms of tree level  Feynman graphs.
External legs of the graphs 
correspond to the wavefunctions of asymptotic states.
In the usual example of QFT in flat space,
these are plane waves.
In the expansion we are setting up in AdS, the
 external legs of graphs are associated with the boundary behavior of $\phi$ (`bulk-to-boundary propagators').
 These diagrams are sometimes called `Witten diagrams,' after \cite{Witten:1998qj}.
\marginpar{\rotatebox{270}{\tiny [Francesco D'Eramo]}}
\begin{figure}[ht]
\begin{center}
\leavevmode\hbox{\epsfxsize=10cm \epsffile{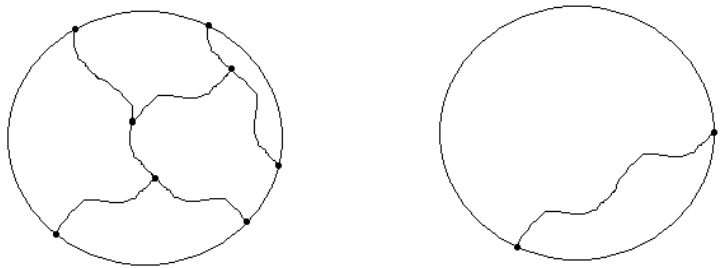}}\\[3mm]
\caption{Feynman graphs in AdS. We did the one with two external legs 
in section \ref{sec:bulktoboundary}.}
\end{center} 
\end{figure}

\subsection{$n$-point functions}

Next we will briefly talk about connected correlation functions of three or more operators.
Unlike two-point functions, such observables are sensitive to the details of the bulk interactions,
and we need to make choices.
For definiteness, we will consider the three point functions of 
scalar operators dual to scalar fields with the following bulk action:
\be
 S = {1 \over 2} \int d^{d+1} x \sqrt{g}  \[ \sum_{i=1}^{3}\( (\p \phi_i)^2 + m_i^2 \phi_i^2\)  + b \phi_1 \phi_2 \phi_3\]~.
\ee
The discussion can easily be extended to $n$-point functions with $n>3$.

The equations of motion are
\be
 (\Box-m_1^2) \phi_1 (z,x) = b \phi_2 \phi_3 
\ee
and its permutations.
We solve this perturbatively in the sources, $\phi_0^i$:
\begin{eqnarray}
\label{perturbativesolution}
&& \underline{\phi}^1 (z,x)=\int d^d x_1 K^{\Delta_1} (z,x;x_1) \phi_0^1 (x_1) \\
 \nonumber
&+&b \int d^d x' dz' \sqrt{g} G^{\Delta_1} (z,x;z',x') \int d^d x_1 \int d^d x_2 
 K^{\Delta_2} (z',x';x_1)  \phi_0^2 (x_1) K^{\Delta_3} (z',x';x_2) \phi_0^3 (x_2) \\
 \nonumber
 &+&\CO(b^2 \phi_0^3)~,
\end{eqnarray}
with similar expressions for $\underline{\phi}^{2,3}$.
We need to define the quantities $K, G$ appearing in (\ref{perturbativesolution}).
$K^\Delta$ is the bulk-to-boundary propagator for a bulk field 
dual to an operator of dimension $\Delta$. We determined this in the previous subsection: 
it is defined to be the solution to the homogeneous scalar wave equation 
$(\Box -m^2)K^{\Delta} (z,x;x') =0$ which approaches a delta function 
at the boundary\footnote{
More precisely, as we saw in the previous subsection, it is better to use a regulated 
bulk-to-boundary propagator which approaches a delta function at the regulated
boundary:
$ K_\epsilon^{\Delta} (\epsilon,x;x')= \epsilon^{\Delta_-}\delta(x, x') $.
}, $ K^{\Delta} (z,x;x') \buildrel{z \to 0} \over{\to} z^{\Delta_-}\delta(x, x') $.
So $K$ is given by equation (\ref{bulktoboundary}) with 
$\phi_0(k) = e^{ - i kx'}$.
$G^{\Delta_i} (z,x;z'x')$ is the bulk-to-bulk propagator, 
which is the {\it normalizable} solution to the wave equation with a source
\be
 (\Box -m_i^2)G^{\Delta_i} (z,x;z',x') = {1 \over \sqrt{g}} \delta (z-z') \delta^d (x-x') 
\ee
(so that $ (\Box -m_i^2) \int \sqrt{g} GJ = J $ for a source $J$).
\marginpar{\rotatebox{270}{\tiny [Daniel Park]}}
\begin{figure}[h]
\hskip-.1in
\begin{minipage}[t]{0.33\linewidth}
 \centering \includegraphics[width=3cm]{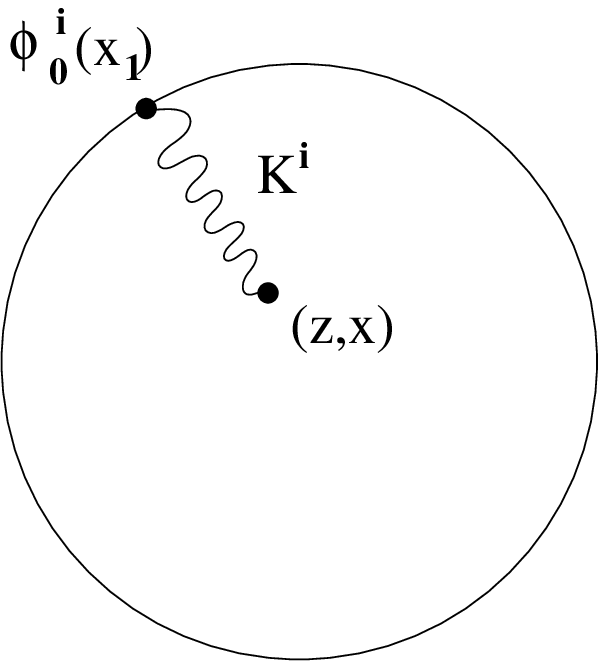}
 \caption{\label{f1}
 Witten diagram 1} 
\end{minipage}%
\hspace{0.4cm}%
\begin{minipage}[t]{0.33\linewidth}
 \centering \includegraphics[width=3cm]{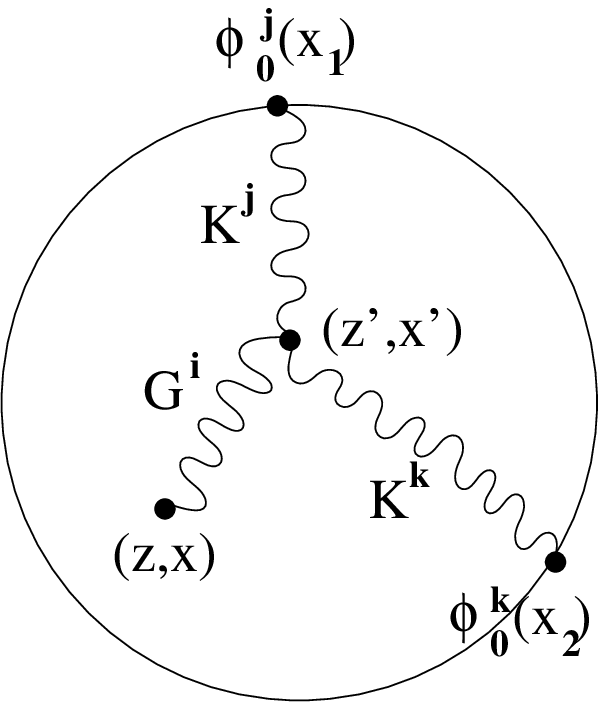}
 \caption{ \label{f2}
 Witten diagram 2}
\end{minipage}%
\hspace{0.4cm}%
\begin{minipage}[t]{0.33\linewidth}
 \centering \includegraphics[width=3cm]{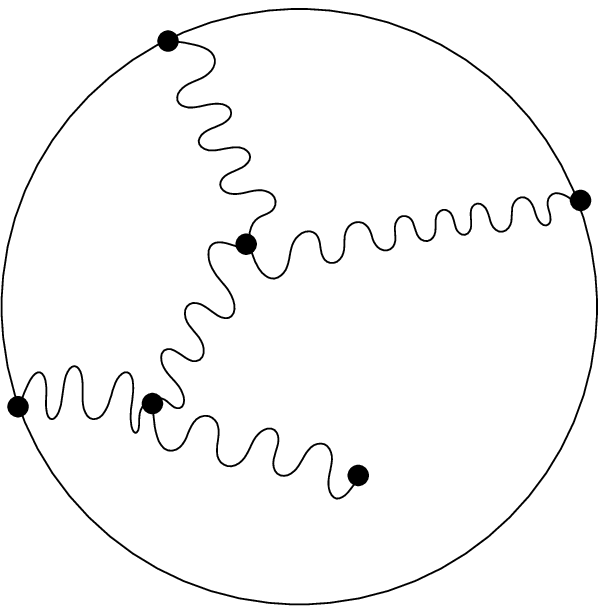}
 \caption{ \label{f3}
 Witten diagram 3}
\end{minipage}
\end{figure}

The first and second terms in (\ref{perturbativesolution}) are summarized by the Witten diagrams in figure \ref{f1} and \ref{f2}.
A typical higher-order diagram would look something like figure \ref{f3}.
This result can be inserted into our master formula (\ref{response})
to find $G_3$.

\subsection{Which scaling dimensions are attainable?}\label{sec:KW}

Let's think about which $\Delta$ are attainable by varying $m$\footnote{In this subsection, we work in units of the AdS radius, \ie\ $L=1$.}.
$\Delta_+ \geq d/2$ is the smallest dimension we've obtained so far,
but the bound from unitarity is lower: $\Delta_+ \geq {d-2 \over 2}$.
There is a range of values for which we have a choice about 
which of $z^{\Delta_\pm}$ is the source and which is the response.
This comes about as follows.

$\phi$ is normalizable if $S[\phi] <\infty$.
With the action we've been using 
\be\label{usualaction}
 S_{Bulk} = \int_\epsilon d^{d+1} x \sqrt{g} (g^{AB} \p_A \phi \p_B \phi +m^2 \phi^2)
\ee
with $\sqrt g = z^{-d-1}$, our boundary conditions demand that
$\phi \sim z^{\Delta}(1+\CO(z^2))$ with
$\Delta=\Delta_+$ or $\Delta_-$,
\be
 g^{zz} (\p_z \phi)^2 = (z \p_z \phi)^2 \sim \Delta^2 z^{2\Delta}
\ee
and hence,
\be
 g^{AB} \p_A \phi \p_B \phi +m^2 \phi^2  \simeq
 \Delta^2 z^{2\Delta} +k^2 z^{2\Delta+2} + m^2 z^2 = (\Delta^2 +m^2) z^{2 \Delta} (1+\CO(z^2))
\ee
in the limit $z\rightarrow 0$. Since for $\Delta =\Delta_\pm$, $\Delta^2 +m^2  = -d \Delta \neq 0$,
\be
 S_{Bulk} [z^\Delta] \sim \int_\epsilon dz z^{-d-1} (-d\Delta) z^{2\Delta} (1+ \CO(z^2))
 \propto {1 \over 2\Delta -d} \epsilon^{2\Delta -d}~.
\ee
We emphasize that we have only specified the boundary behavior of $\phi$,
and it is not assumed that $\phi$ satisfies the equation of motion.
We see that
\be
S_{Bulk} [z^\Delta] < \infty~~~ \Leftrightarrow ~~~\Delta > d/2.
\ee
This does not saturate the lower bound 
from unitarity on the dimension of a scalar
operator, which is $\Delta > {d-2 \over 2}$;
the bound coincides with the engineering dimension of a free field.


We can change which fall-off is the source by adding
a boundary term to the action (\ref{usualaction}) \cite{Klebanov:1999tb}:
\be
 S^{KW}_{Bulk} = \int_\epsilon d^{d+1} x \sqrt{g} \phi(-\Box +m^2 )\phi
 = S_{Bulk}- \int_{\p AdS} \sqrt{\gamma} \phi n \cdot \p_z \phi~~.
\ee
For this action we see that
\begin{align*}
\begin{split}
 S^{KW}_{Bulk}[\phi \sim z^{\Delta} (1+\CO(z^2))]
 &= \int_\epsilon dz z^{-d-1} z^{\Delta} (1+\CO(z^2)) [(-\Delta(\Delta-d)+m^2)z^{\Delta}(1+\CO(z^2))+k^2 z^{2\Delta+2}] \\
 &\sim \int_\epsilon dz z^{-d-1+2\Delta+2} \sim \epsilon ^{2\Delta -d+2} < \infty
\end{split}
\end{align*}
is equivalent to
\be
 \Delta \geq {d-2 \over 2}
\ee
which is exactly the unitary bound.
We see that in this case both $\Delta_\pm$ give normalizable
modes for $\nu \leq 1$.
Note that it is actually $\Delta_-$ that gives the value which saturates the unitarity bound,
that is, when
\be
\Delta_- = \( {d \over 2} - \sqrt{\({d \over 2}\)^2 +m^2} \)_{m^2=1- {d^2 \over 4}} = {d-2 \over 2}~~.
\ee
The coefficient of $z^{\Delta_+}$ would be the source in this case.

We have found a description of a different boundary CFT from the same bulk action,
which we have obtained by adding a boundary term to the action.
The effect of the new boundary term is to lead us to impose Neumann boundary conditions
on $\phi$, rather than Dirichlet conditions.
The procedure of interchanging the two is similar
to a Legendre transformation.

\subsection{Geometric optics limit}

When the dimension of our operator $\CO$ is very large, 
the mass of the bulk field is large in units of the $AdS$ radius:
\be m^2 L^2 = \Delta (\Delta - d) \sim \Delta^2 \gg 1 ~~~\Longrightarrow~~~ mL \gg 1. \ee
This means that the path integral which
computes the solution of the bulk wave equation
has a sharply peaked saddle point, associated with 
geodesics in the bulk.
That is, we can think of the solution of the bulk wave equation
from a first-quantized point of view, in terms of particles of mass $m$ in the bulk.
For convenience, consider the case of a complex operator $\CO$, so that the worldlines
of these particles are oriented.
Then 
\begin{equation}
\vev{\CO(+a)\bar{\CO}(-a)} = Z[\pm a] \buildrel{m\gg L^{-1}}\over\sim \exp(-S[\underbar{z}]) .
\end{equation}
($\bar \CO$ is the complex conjugate operator.)
The middle expression is
the Feynman path integral $Z[\pm a]=\int [dz(\tau)] \exp(-S[z]  
)$;
the action for a point particle of mass $m$ whose world-line is $\Sigma$ is given by $S[z] = m \int_\Sigma ds$. 
In the limit of large $m$, we have
\begin{equation}
Z[\pm a] \sim \exp (-S[\underbar{z}])
\end{equation}
where we have used the saddle point approximation; $\underbar{z}$ is the geodesic connecting the points $\pm a$ on the boundary.

\marginpar{
\rotatebox{270}
{\tiny [Vijay Kumar]}}
\begin{figure}[h]
\centering
\input{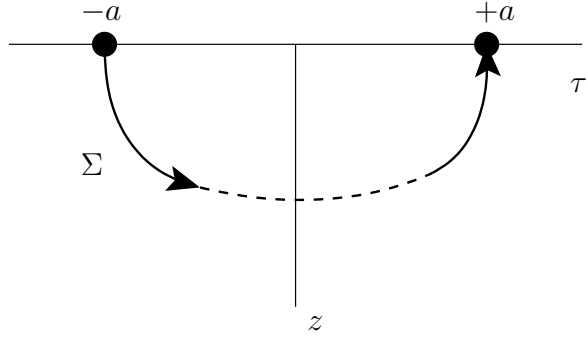}
\caption{ \label{fig1}
The curve $\Sigma \subset \ads$ connecting the two points in $C$. The arrows on the curve indicate the orientation.}
\end{figure}
We now compute $Z[\pm a]$ in the saddle point approximation. The metric restricted to $\Sigma$ is given by $ds^2 \vert_\Sigma = \frac{L^2}{z^2} (1+z'^2) d\tau^2$, where $z':=dz/d\tau$. This implies that the action is
\begin{equation}
S[z] = \int d\tau \frac{L}{z} \sqrt{1+z'^2} ~.
\end{equation}
The geodesic can be computed by noting that the action $S[z]$ does not depend on $\tau$ explicitly. This implies we have a conserved quantity 
\begin{eqnarray}
h & = & z'\frac{\partial \CL}{\partial z'}-\CL = \frac{L}{z}\frac{1}{\sqrt{1+z'^2}} \\
\Rightarrow z'^2 & = & \left( \frac{L^2}{h^2} - z^2 \right) \frac{1}{z^2}~~.
\end{eqnarray}
The above is a first order differential equation with solution $\tau = \sqrt{z_{max}^2-z^2}$, with $z(\tau=0)=z_{max}=L/h=a$. This is the equation of a semi-circle. Substituting the solution back into the action gives 
\begin{eqnarray}
S[\underbar{z}] & = & \int d\tau \frac{L}{\underbar{z}} \sqrt{1+\underbar{z}'^2} = 2L \int_\epsilon^a \frac{dz}{\sqrt{a^2-\underbar z^2}} \frac{a}{\underbar z} \\
& = & 2L \log \frac{2a}{\epsilon} + (\mbox{terms } \rightarrow 0 \mbox { as } \epsilon \rightarrow 0)~~.
\end{eqnarray}
You might think that since we are computing this in a conformal field theory, the only scale in the problem is $a$ and therefore the path integral should be independent of $a$. This argument fails in the case at hand because there are two scales: $a$ and $\epsilon$, the UV cutoff. The scale transformation is anomalous and this is manifested in the $\epsilon$ dependence of $S[\underbar{z}]$:
\begin{equation}
\vev{\CO(+a)\bar{\CO}(-a)} = Z[\pm a] \buildrel{m\gg L^{-1}}\over\sim \exp(-S[\underbar{z}]) 
\sim \frac{1}{a^{2mL}} ~~.
\ee
This is exactly what we expect for the two point function, since $m^2L^2 = \Delta(\Delta -d) \approx \Delta^2$ in the large $\Delta$ limit. Without this anomaly, the two point function of the operator $\CO$ would be independent of $a$ forcing $\Delta = 0$, which is impossible in a unitary CFT. This is similar to the anomaly that gives a non-zero scaling dimension of $k^2/4$ to the operator $\exp (ikX)$ for the 2d free boson CFT which has classical scaling dimension zero. Graham and Witten \cite{grawit} showed that such a scale anomaly exists for surface observables for any even $k$. This relation between correlators of large $\Delta$ observables and geodesics in the bulk theory offers a probe of the bulk spacetime \cite{shenker, Kraus:2002iv}.

\subsection{Comment on the physics of the warp factor}

Recall that the bulk radial coordinate behaves like a spectrograph separating the theory into energy scales.  
$z = 0$ is associated with the UV while $z \rightarrow \infty$ describes the IR. 

One way to think about the implementation of this relationship between the radial coordinate
of the bulk geometry and the field theory energy scale
is as follows.
The geometry dual to a more general Lorentz-invariant state is
\be\label{warping} 
ds^2 =  w^2(z) \(- dt^2 + d\vec x^2+ dz^2 \)  ,\ee
where $w(z)$ is called the `warp factor'.
The warp factor for AdS is simply $L/z$.
The coordinates $t, \vec x$ in (\ref{warping}) 
are the field theory time and space coordinates.  
This means that the size of a field theory object 
is related to the the size of its holographic image by the relation:
$$ ({\rm size})_{FT}  = { 1\over w(z) } \( {\rm proper~size} \)  $$
and similarly the energy of a field theory object 
is related to the proper energy of the corresponding bulk excitation by
$$ E_{FT} \sim i \partial_t = w(z) E_{{\rm proper}} .$$

The fact that AdS goes forever in the z direction is related to the fact that the dual theory is a conformal field theory with degrees of freedom at all energies.  
We can interpret the existence of modes at arbitrarily low energy as the statement that the warp factor has a zero at $z = \infty$.  More precisely, there are $\CO(\hl{N^2}))$ degrees of freedom at every energy scale in the CFT.
One concrete manifestation of this is the fact
that we found a continuum of solutions of $(- \Box + m^2 ) \phi = 0 $ --
in particular, for any $k^\mu$ we could match the
required boundary condition in the IR region, $z \to \infty$.

In the gravity dual of a QFT with an energy gap, the warp factor 
has a {\it minimum}.  
This will impose a boundary condition on the bulk wave equation 
at the IR end of the geometry.
The problem becomes like QM of a 1d particle trapped at both ends, and therefore has a discrete spectrum.

There exist gravity solutions 
\cite{Witten:1998zw, Klebanov:2000hb, Maldacena:2000yy}
which are dual to field theories
with logarithmically running couplings which become strong
in the IR, leading to a gap to almost all excitations (except
some order $\hl{N}^0$ number of Goldstone bosons), as in QCD.
%


\section{Finite temperature and density}

$AdS$ is scale invariant.  It's a solution dual to the {\it vacuum} of a CFT.
The correspondence we've developed can also describe systems
which are not scale invariant.  
This includes QFTs with a scale in the lagrangian, or where a scale is generated 
quantum mechanically.
A simpler thing to describe is the dual 
of a CFT in an ensemble which introduces a scale.  
A saddle point of the bulk path integral describing such a state 
would correspond to a geometry which approaches $AdS$ near the boundary,
and solves the same equations of motion
$$ 0 = R_{AB} +{d \over L^2} g_{AB} \propto { \delta S_{{\rm bulk}} \over \delta g^{AB}},$$
 but does something else in the interior.

\subsection{Interjection on expectations for CFT at finite temperature}
In particular we mean a $d$ dimensional relativistic CFT. The partition function is 
\begin{equation}
	Z(\tau) = \text{tr} e^{-H/T} = e^{-F/T}
\end{equation} with free energy $F$, on a space with geometry $S^1_{th} \times \Sigma_{d-1}$ where the $S_{th}^1$ has radius $1/T, \tau \sim \tau + 1/T$ and $\Sigma_{d-1}$ is some $(d-1)$-manifold. We can give $\Sigma_{d-1}$ finite volume as an IR regulator. 
The temperature is a deformation of the IR physics (modes with $\omega \gg T = E_{KK}$ don't notice).

For large $V= {\rm Vol}(\Sigma_{d-1})$, then 
\begin{equation}\label{freeenergy}
F = -c V T^d
\end{equation}
which 
follows from extensivity of $F$ and dimensional analysis.  This is the familiar Stefan-Boltzmann behavior
for blackbody radiation.
The pressure is $P = - \partial_V F$.
Note that in a relativistic theory, just putting it at finite temperature
is enough to cause stuff to be present, because of the existence
of antiparticles.  It is the physics of this collection of CFT stuff
that we would like to understand.

The prefactor $c$ in (\ref{freeenergy}) is a measure 
the number of degrees-of-freedom-per-site, \ie\ the number of species
of fields in the CFT, which we called `$\hl{N^2}$' above.

\subsection{Back to the gravity dual}

The desired object goes by many names, such as planar black hole, Poincar\'e black hole, black brane... Let's just call it a black hole in $AdS_{d+1}$.
The metric is
\begin{eqnarray}\label{AdSBH}
	ds^2 &=& \frac{L^2}{z^2} \left( -f dt^2 + d\vec x^2 + \frac{dz^2}{f} \right) \nn \\
	f &=& 1 - \frac{z^d}{z_H^d}~~.
\end{eqnarray} We again put the $\vec x$ coordinates on a finite volume space, for example in box of volume $V, x \sim x + V^{1/d}$. Notice that 
if we set the emblackening factor $f=1$ we recover the Poincar\'e AdS metric, and in fact $f$ 
approaches $1$ as $z \to 0$, demonstrating that this is an IR deformation.

This metric solves Einstein's equations with a cosmological constant $\Lambda = -\frac{d(d-1)}{2L^2}$ and asymptotes to Poincar\'e AdS.
It has a horizon at $z=z_H$, where the emblackening factor $f$ vanishes linearly.
This means that events at $z > z_H$ can't influence the boundary near $z=0$.

The fact that the horizon is actually translation invariant
(\ie\ it's a copy of $\IR^{d-1}$, rather than a sphere) leads some people to call this a `black brane'.

In general, horizons describe thermally mixed states.
Here we can see the connection more directly: 
this solution (\ref{AdSBH}) is the extremum of the euclidean gravity action
dual to the QFT path integral with thermal boundary conditions.
Recall that the boundary behavior of the bulk metric
acts as a source for the boundary stress tensor -- 
changing the boundary behavior of the bulk metric 
is the same as changing the metric of the space
on which we put the boundary theory.
This is to say that if
$$ ds^2_{bulk} \buildrel{z \to 0 }\over \approx {dz^2 \over z^2} + {L^2 \over z^2} 
g_{\mu\nu}^{(0)} dx^\mu dx^\nu $$
then, up to a factor, the boundary metric is $g_{\mu\nu}^{(0)}$.
This includes any periodic identifications we might make on the geometry,
such as making the euclidean time periodic.

A study of the geometry near the horizon 
will give a relationship between the temperature and 
the parameter $z_H$ appearing in the metric (\ref{AdSBH}).
The near-horizon ($z \sim z_H$) metric is
\begin{equation}
	ds^2 \sim - \kappa^2 \rho^2 dt^2 + d\rho^2 + \frac{L^2}{z_H^2}d\vec x^2
\end{equation} 
where $ \rho^2 = {2  \over \kappa}{L\over z_H^2} (z_H-z)  + \CO(z_H-z)^2$, 
and $\kappa \equiv  {|f'(z_H)|/2} = {d/2 z_H}$ 
is called the `surface gravity'.
If we analytically continue this geometry to euclidean time, $ t \to i \tau$, 
it becomes
\begin{equation}
	ds^2 \sim \kappa^2 \rho^2 d\tau^2 + d\rho^2 + \frac{L^2}{z_H^2}d\vec x^2
\end{equation} 
which looks like $\IR^{d-1}$ times a euclidean plane in 
polar coordinates $\rho, \kappa \tau$.
There is a deficit angle in this plane unless $\tau$ is periodic according to
$$ \kappa \tau \simeq \kappa \tau + 2\pi .$$
A deficit angle would mean nonzero Ricci scalar curvature, which would mean that 
the geometry is {\it not} a saddle point of our bulk path integral.
Therefore we identify the temperature as 
$T = \kappa/(2\pi) = d/(4 \pi z_H)$. 

Meanwhile the area of the horizon (the set of points with $z=z_H, t$ fixed) is
\begin{equation}
	A = \int_{z=z_H, \text{fixed} t} \sqrt{g} d^{d-1} x = \left( \frac{L}{z_H} \right)^{d-1} V~~.
\end{equation} Therefore the 
entropy is 
\begin{equation}
	S = \frac{A}{4 G_N} = \frac{L^{d-1}}{4 G_N} \frac{V}{z_H^{d-1}} = \frac{N^2}{2\pi} (\pi T)^{d-1}V = \frac{\pi^2}{2} N^2 V T^{d-1}~~.
\end{equation} 
Here I have used the relation $\frac{L^{d-1}}{4 G_N} = \frac{N^2}{2\pi}$.
(The factor of $2\pi$ is the correct factor for the case of the duality involving the $\CN=4$ SYM, so this factor of $N^2$ is not red.)
The entropy {\it density} is 
$$s_{BH} = { S_{BH} \over V} = {a_{BH} \over 4 G_N} .$$
where $a_{BH} \equiv {A \over V}$ is the `area density' (area per unit volume!) of the black hole.

We would like to identify this entropy with that of the CFT 
at finite temperature.
The power of $T$ is certainly consistent with the constraints of scale invariance.
It is not a coincidence that the number of degrees-of-freedom-per-site
$ \frac{L^{d-1}}{G_N}$ appears in the prefactor.  
In the concrete example of $\CN=4$ SYM in $d=3+1$, we can compute 
this prefactor in the weak-coupling limit $\lambda \to 0 $
and the answer is \cite{Gubser:1996de}
$$ F(\lambda = 0) = {3\over 4}  F(\lambda = \infty) ~;$$
the effect of strong coupling is to reduce the effective  
number of degrees-of-freedom-per-site by an order-one factor.
Similar behavior is also seen in lattice simulations of QCD \cite{latticeref}.

To support the claim that this metric describes 
the saddle point of the partition sum of 
a CFT in thermal equilibrium,
consider again the partition function:
$$Z_{CFT}\equiv e^{-\beta F}= e^{-S_g[{\underline g}] } ,$$ 
where $\underline{g}$ is the euclidean saddle-point metric\footnote{If there were more than one saddle point geometry
with the required asymptotics, we would need to sum
over them.  
In fact there are examples where there are multiple saddle points
which even have different topology in the bulk, which do exchange dominance
as a function of temperature
\cite{Hawking:1982dh, Witten:1998zw}.
In this example, this behavior matches a known phase transition in the dual gauge theory.
This is therefore strong evidence that quantum gravity involves
summing over topologies.}.
$S_g$ is the on-shell gravity action for the black hole solution and is equal to
\begin{equation}
	S_g = S_{EH} + S_{GH} + S_{ct} ~.
\end{equation} 
$$S_{EH} = - { 1\over 16 \pi G_N} \int d^{d+1} x \sqrt g \left( 
R + {d (d-1) \over L^2} \right) $$
is just the usual bulk gravity action.
In addition to this there are two boundary terms.
The `Gibbons-Hawking' term is an extrinsic boundary term
which affects which boundary conditions we impose on the metric,
in the same way that the $ \int_{\partial AdS} \phi n \cdot \partial \phi $
term in the scalar case changed the scalar boundary conditions from
Neumann to Dirichlet.
Its role is to guarantee that when we impose Dirichlet boundary conditions 
on the metric by specifying $g_{\mu\nu}^{(0)}$, 
the action evaluated on a solution is stationary under an arbitrary
variation of the metric satisfying that boundary condition\footnote{Without this term, integration by parts in the Einstein-Hilbert term
to get the EOM produces some boundary terms proportional
to variations of derivatives of the metric:
$$\delta S_{EH} = EOM + \int_{\partial AdS} \gamma^{\mu\nu} n \cdot  
\partial \delta \gamma_{\mu\nu} ,$$
which is incompatible with imposing a Dirichlet condition on the metric.
}.  Its specific form is:
$$S_{GH} = - 2 \int_{\partial M} d^d x \sqrt{\gamma} \Theta $$
where $\Theta $ is the extrinsic curvature of the boundary
$$ \Theta \equiv \gamma^{\mu\nu} \nabla_\mu n_\nu = 
{ n^z \over 2} \gamma^{\mu\nu} \partial_z \gamma_{\mu\nu} .$$
where $n^A$ is an outward-pointing unit normal to the boundary $z=\epsilon$,
and we've defined $\gamma$ by 
$$ds^2 \buildrel{z\to0}\over \approx L^2 {dz^2\over z^2} + \gamma_{\mu\nu} dx^\mu dx^\nu .
$$

Finally, 
$$S_{ct} = \int_{\partial M} d^d x \sqrt{\gamma} { 2(d-1) \over L} + \dots $$
is a local, intrinsic boundary counter-term, as we need to subtract some divergences as $z \to 0$, just like in the calculation of vacuum correlation functions of local operators.
The $\dots$ are terms proportional to the intrinsic curvature of the boundary metric 
$g_{\mu\nu}^{(0)}$, which we have taken to be flat.
See \cite{Henningson:1998gx, Balasubramanian:1999re} for more details.

Thus by plugging in the AdS planar black hole solution (the saddle point) we obtain the free energy.
Specializing to $d=3+1$ and using the $\CN=4$ SYM normalization for the rest of this section, we find
\begin{equation}
	-\frac{F}{V} = \frac{L^2}{16 \pi G_N} \frac{1}{z_H^4} = \frac{\pi^2}{8} N^2 T^4.
\end{equation}
This is consistent with the Bekenstein-Hawking entropy calculation above:
$$ S_{BH} = -\partial_T F. $$
We can also check the first law of thermodynamics:
\be\label{firstlaw} dE + PdV = T_{BH} dS_{BH} \hl{+ \Omega dJ + \Phi dQ}, \ee
where the pressure is $P = - \partial_V F$.
This is actually a strong check (at least on the correctness of our numerical factors), because it relates horizon quantities such as $T,S$ to global quantities such as the free energy $F$. 
The red terms on the RHS of (\ref{firstlaw}) 
would be present if we were studying a black hole
with angular momentum $J$, or with charge $Q$ as in the 
next subsection.

We can also compute the expectation value of the field theory stress tensor $T^{\mu \nu}$. We 
just use the usual GKPW prescription, using the fact that $T^{\mu\nu}$ couples to the induced metric on the boundary $\gamma_{\mu \nu}$. The energy is $E = -V \sqrt{\gamma} T^t_t$ and the pressure $P =V \sqrt{\gamma} T^x_x$.
The answers agree with the form we found from thermodynamics:
\be (E/V) = {N^2\pi^2\over 8} {3\over 2 z_H^4},~~~~~~~ P = {N^2\pi^2\over 8} {1 \over 2z_H^4}\ee 
which satisfies $E = 3P$, so $T^\mu_\mu = 0$ as required in a 3+1-dimensional relativistic CFT. 
One can do the same for 
the black hole with a spherical horizon (in `global coordinates')
which describes the field theory on a sphere. 
The answer is $E ={N^2\pi^2\over 8} \( 3/(2 z_H^4) + 3L^2/(8G)\) $. 
The limit $z_H \to \infty$ describes 
the zero-temperature vacuum.
In this limit $E$ is nonzero, and in fact matches the calculation for the zero-point (Casimir) energy 
of the field theory on a sphere of radius $L$.
The fact that this quantity matches precisely the field theory result at weak coupling \cite{Henningson:1998gx, Balasubramanian:1999re} (unlike, for example, the free energy which depends on the 't Hooft coupling) is because it arises from an anomaly.

\subsection{Finite density}
Suppose the boundary theory has a conserved $U(1)$ symmetry,
with current $J^\mu$.
This means that there should be a massless gauge field $A_\mu$ in the bulk.
Naturalness suggests the action
\be \Delta S_{{\rm bulk}} = - {1\over 4 g_F^2} \int d^{d+1} x \sqrt g F_{\mu\nu} F^{\mu\nu} , ~~~~
F_{\mu\nu} \equiv \partial_\mu A_\nu-\partial_\nu A_\mu. \ee
The Maxwell equation near the boundary implies the behavior
\be A \buildrel {z\to0}\over \approx A^{(0)}(x) + z^{d-2}A^{(1)}(x), \ee
and in particular for the zeromode of the time component,
$A_t  \buildrel {z\to0}\over \approx  \mu + z^{d-2} \rho $.
Using the result (\ref{response}), 
this is the statement that the charge density $\rho$ of the boundary
theory is related to the electric field in the bulk:
\be \Pi_{A_t} = {\partial \CL \over \partial (\partial_z A_t)} 
= E_z = A^{(1)} = \rho ~; \ee
the subleading behavior encodes the field momentum.

A black hole which describes the equilibrium configuration
of the field theory at finite density has the same form of the metric as above
\be	ds^2 = \frac{L^2}{z^2} \left( -f dt^2 + d\vec x^2 + \frac{dz^2}{f} \right)
\ee
but with a different emblackening factor $f = 1 - M z^d + Q z^{2d-2} $, 
and a nonzero gauge field
\be A= dt \( \mu + \rho z^{d-2} \) .\ee
$M, Q, \rho$ can be written in terms of $g_F, \mu, T$, see \eg\ 
\cite{Hartnoll:2009sz}.

I will restrict myself to two comments about this solution.
First, in the grand canonical ensemble, $\mu$ is fixed,
and we should think of the $z^0$ term in $A_t$ as the source,
and the coefficient of $z^{d-2}$, namely $\rho$ as the response.
Changing the boundary conditions on the gauge field 
can be accomplished as described in (\ref{sec:KW}), 
by adding a term of the form $ \int_{\partial AdS} n_\mu A_\nu F^{\mu\nu} $.
In this case, the trick {\it is} the Legendre transformation which
takes us to the canonical ensemble,
where the $\rho$-term is the source and $\mu$ is the response.

Secondly, this geometry is interesting.
At $T \ll \mu$ it describes an example of a `holographic RG flow' between
RG fixed points.
In particular, the function $f$ now has multiple zeros
at real, positive values of $z$.
Let's call the one closest to the boundary $z_0$;
this represents a horizon;
there is another zero $z_1$ at a larger value of $z$,
in the inaccessible region behind the horizon.
These zeros collide $z_1 \to z_0$ when we take $T \to 0$,
in which case $f$ has a double-zero, and the black hole 
is said to be extremal.  
In this limit, the geometry interpolates between $AdS_{d+1}$ near the boundary
$z \sim 0$, 
and $AdS_2 \times \IR^{d-1}$ near the horizon.
The presence of an $AdS$ factor means that this IR region of the geometry is scale invariant.  The original scale invariance of the $d$-dimensional CFT is broken
by the chemical potential, so this is an {\it emergent} scale invariance,
like the kind that one might see in a real system.  
The IR geometry is dual to some non-relativistic scale-invariant theory.
At finite $T \ll \mu$, the IR geometry is a black hole in $AdS_2 \times \IR^{d-1}$,
related to this CFT at finite temperature.
\begin{figure}[h]
\begin{center}
\includegraphics[scale=0.60]{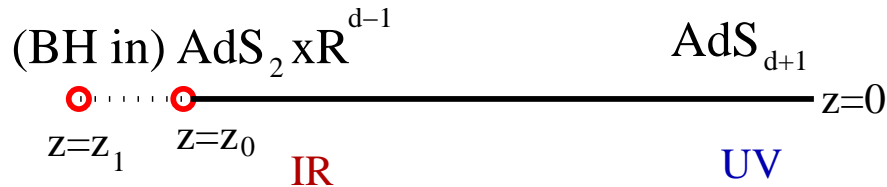}
\end{center}
\caption{\label{extremalBH}
The geometry of the near-extremal $AdS_{d+1}$ charged black hole 
for $T \ll \mu$.}
\end{figure}

If we put a charged boson in this geometry, we find the holographic superconductor
phenomenon (\cite{Gubser:2008px,Hartnoll:2008vx,Hartnoll:2008kx}, 
for a review see \cite{Herzog:2009xv}).
If we put a charged spinor in this geometry, we can study the 
two point functions of the dual fermionic operator, 
and in particular its spectral density $\sim \Im G^R$.
This study was initiated in \cite{Lee:2008xf} 
and developed by \cite{LMV, FLMV, Cubrovic:2009ye},
and one finds Fermi surfaces, which sometimes
describe non-Fermi liquids.
For some values of the bulk parameters,
the fermion spectral density is that of a 
`marginal Fermi liquid' introduced \cite{varma}
to model the strange metal phase of 
high-$T_c$ superconductors \cite{FLMV}.

\section{Hydrodynamics and response functions}

So far we have given evidence that the black hole thermodynamics
of the AdS black hole solution is dual 
to the thermal ensemble of some strongly coupled CFT on $\IR^{d-1}$ at
large $N$ and large 't Hooft coupling $\lambda$. 
Thus the gauge theory provides the
microstates that are being coarse grained by the Bekenstein-Hawking
entropy of the black hole $S_{BH}$.
The static black hole describes the 
field theory in thermodynamic equilibrium.

A few comments on this observation:

\begin{itemize}
\item Perturbing the equilibrium of the \emph{boundary} theory with
a kick will result in thermalization - relaxation
back to equilibrium. 

\item In the \emph{bulk} the response to such a kick is for
the energy of the kick to fall into the black hole.  
\end{itemize}

The above two statements are related by the duality \cite{Horowitz:1999jd}.
In the long wavelength and small frequency limit both are
consistent with the hydrodynamics of a relativistic CFT.
Additionally the duality allows one to compute various
transport coefficients of the gauge theory at large $\lambda$,
such as the shear viscosity, conductivity, etc
as we discuss next.
We emphasize that these calculations are done at leading order 
in this mean-field-theory-like description, but they include
dissipation.

\subsection{Linear response, transport coefficients}

As a demonstration of the real-time prescription \cite{Son:2002sd}, 
in the following subsection we will derive the viscosity of a large class of strongly-interacting plasmas,
made from CFT at finite temperature. 
First, we quickly recall the formalism of linear response to establish notation.

To see how our system responds to our poking at it,
consider the following small perturbation of the field theory action
\be \Delta S_{QFT} = \int d^d x \CO \phi_0, ~~\phi_0~{\rm small} ~~.\ee
The response is
\be \delta \vev{\CO}_{CFT, T} \buildrel{\phi_0 \to 0 }\over = 
- G^R (\omega, {\bf k}) \phi_0(\omega, {\bf k}). 
\ee
For simplicity, we are just asking about the diagonal response,
perturbing with the operator $\CO$, and measuring the operator $\CO$:
$ G^R \equiv G^R_{\CO \CO} $.
The subscript on the LHS indicates that we are computing thermal
averages at temperature $T$.
In the long-wavelength, low-frequency limit, 
on very general grounds, this will reduce to the Kubo formula
\be\label{kubo} 
 \delta \vev{\CO}_{CFT, T} \buildrel{k  \to 0, \omega \to 0 }\over \to i \omega \chi \phi_0 .\ee
For example, in the case where $\CO = j^\mu$ is a conserved current, 
$\phi_0 = A_\mu $ is the boundary behavior of a bulk gauge field,
and the transport coefficient is the conductivity:
$$  \delta \vev{\vec J}_{CFT, T} \buildrel{k  \to 0, \omega \to 0 }\over \to i \omega \chi({j}) \vec A  
= \sigma \vec E~~~~({\rm Ohm's~law}).$$
In the case where $\CO = T^x_y(k_z)$, the source
is the boundary value of a metric perturbation $\delta g^y_x$,
and the transport coefficient is the shear viscosity, $\chi\({T^x_y}\) = \eta$.
Don't forget that the order of limits here matters: $k$ must be taken to zero before $\omega$
to get the transport coefficient.

\subsection{Holographic calculation of transport coefficients}
\label{sec:nabil}

Now we discuss the bulk calculation of these quantities.
We will follow the discussion of \cite{Iqbal:2008by}, but we should emphasize that this calculation has a long history \cite{Buchel:2003tz, Kovtun:2003wp, Kovtun:2004de, 
Starinets:2008fb}.
We can consider a very general bulk metric:
$$ ds^2 = g_{tt} dt^2 + g_{zz} dz^2 + g_{ij} dx^i dx^j $$
which satisfies two conditions:
\begin{itemize}
\item Near $z \to 0$, it approaches $AdS$, or some other asymptotics
for which we know how to construct a holographic GKPW-like prescription.
Other examples include systems which in the UV are described by non-relativistic CFTs,
as described in \cite{Son:2008ye, Balasubramanian:2008dm, Kachru:2008yh}.
\item The geometry has a horizon at some $z=z_H$, near which the metric coefficients
take the form $g_{tt} \approx  2 \kappa (z_H -z), g_{zz} \approx { 1\over 2 \kappa (z_H - z) } .$  Such a thing is called a Rindler horizon,
and means that the euclidean time coordinate must have period $ {1\over T} = {2 \pi \over \kappa}. $
$T$ is therefore the temperature at which we've put the field theory.
\end{itemize}

In this spacetime, consider the bulk action
$$ S = - \half \int d^{d+1} x \sqrt g {  \partial_A \phi\partial^A \phi \over q(z) } .$$
Several points are noteworthy here.
First, the quantity $q(z)$ is some effective coupling of the mode, 
which can depend on the radial direction; such dependence can arise
for example from the profiles of some background fields to which the mode is coupled.
In addition, we've assumed that the field $\phi$ is massless, and 
does not mix with other modes (at least in the kinematical regime of interest);
this will be important for the calculation below.
An important example of such a $\phi$ is precisely the metric fluctuation
$\delta g_x^y(k_z)$ which computes the shear viscosity. 
In the case of Einstein gravity, the coefficient is 
${1 \over q_{{\rm Einstein}} } = {1\over 16 \pi G_N} $.
For other theories of gravity with higher-curvature corrections,
$q$ can take some other form \cite{Iqbal:2008by}.
We will continue to leave $q$ general, and refer to the 
operator to which $\phi$ couples as $\CO$.

Recall that the general formula for the expectation value
of an operator in terms of the behavior of its dual bulk field is
\be \label{masslessvev} \vev{\CO(x^\mu) }_{QFT} = \lim_{z\to0} \Pi_\phi (z, x^\mu) \ee
where $\Pi_\phi \equiv { \partial \CL_{{\rm bulk}} \over \partial ( \partial_z \phi) } $
is the field momentum (with the radial direction thought of as time).
We have specialized the formula (\ref{response}) to the case of a massless field.
The fact that (\ref{masslessvev}), evaluated on the infalling solution for $\phi$, correctly computes the retarded response was demonstrated in \cite{Iqbal:2009fd}.
Equation (\ref{masslessvev})  means that the transport coefficient is
\be \chi = \lim_{\omega \to 0} \lim_{k\to 0 }  \lim_{z\to 0} \left( { \Pi_\phi(z, k_\mu) \over 
i \omega \phi(z, k_\mu) } \right). \ee
We will calculate this in two steps.
First we find its value at the black hole horizon, and then we propagate it to 
the boundary using the equation of motion.

By assumption (2) about the metric, the horizon at $z=z_H$ is a 
regular singular point of the wave equation, near which solutions behave as 
\be \phi(z) \approx (z_H- z)^{\pm {i\omega \over 4 \pi T} }
\ee
(the exponents are determined by plugging in a power-law ansatz 
into the wave equation and Taylor expanding).
Since the time-dependence of the solution is of the form $e^{ - i \omega t}$ 
(recall equation (\ref{eq2})),
these two solutions describe waves which fall into (-) 
or come out of (+) the black hole horizon as time passes.  
To compute the retarded Green's function,
we pick the ingoing solution \cite{Son:2002sd}.
This says that near the horizon, the field momentum is
\be
 \Pi_\phi(z, k) = {\sqrt{|g|}\over q(z)} g^{zz} \partial_z \phi 
\buildrel{z\to z_H}\over{\approx} 
{1\over q(z_H)} \sqrt{\left| {g \over g_{zz} g_{tt}} \right|} i \omega \phi(z, k)|_{z=z_H} . 
\ee
The outgoing solution would give a minus sign in front here.

To propagate this to the boundary, we use the bulk equations of motion,
which relate $ \partial_z \Pi_\phi \sim\partial_z^2 \phi$ to $\phi$.
It is not hard to show that in the limit $k\to 0, \omega \to 0$, 
\be\partial_z \Pi_\phi( z, k \to 0, \omega \to0) = 0 .\ee
A similar statement holds for the denominator of the Green's function, $\omega \phi$.
This means that 
\be {\Pi \over \omega \phi} |_{z=0}
={\Pi \over \omega \phi} |_{z=z_H} \ee
from which we learn that 
\be \chi  ={1\over q(z_H)} \sqrt{\left| {g \over g_{zz} g_{tt}} \right|}. \ee
Here it was important that the bulk field was massless;
this fails for example for the mode which computes the bulk viscosity.

Let's apply this discussion to the case where $\chi$ is the shear viscosity $\eta$, defined 
in the previous subsection.
The shear viscosity is dimensionful (it comes in some units called `poise');
a dimensionless measure of the quality of a liquid is 
its ratio with the entropy density, which is also something we know how to compute.
The entropy density of our system is related to the `area-density' $a_{BH}$ of the black hole
\be s = { a_{BH} \over 4 G_{N} } = 
{1\over 4 G_N}  \sqrt{\left| {g \over g_{zz} g_{tt}} \right|} ~~.\ee
We see therefore that 
\be {\eta \over s} = {1\over q(z_H) } . \ee
In the special case of Einstein gravity in the bulk this gives the 
celebrated KSS value \cite{Kovtun:2004de}
\be  {\eta \over s} = {1\over 4\pi}. \ee
This value is much smaller than that of common liquids.
The substances which come the closest 
\cite{Schafer}
are cold atoms at unitarity 
(${\eta \over s } \sim \half$ \cite{ThomasVisc}) and
the fireball at RHIC (${\eta \over s } \sim 0.16 $  \cite{ShuryakEQ}).
This computation of the shear viscosity of a strongly-interacting 
plasma seems to have been quite valuable to people
trying to interpret the experiments at RHIC\footnote{For
reviews of applications of holographic duality to RHIC, please see \cite{qgpreviews}.}.

It seems not to be a lower bound.  
For example, in some particular higher-curvature gravity theory
called Gauss-Bonnet gravity,
where the black-hole solution is known,
the parameter $q$ is related to a coefficient of a higher-derivative term $\lambda_{GB}$,
and one finds \cite{Brigante:2008gz}
\be {\eta \over s} = {1\over q(z_H)} = { 1 - 4 \lambda_{GB} \over 4\pi} > {16 \over 25} .\ee
The inequality on the right arises from demanding causality of the boundary theory,
which fails if $\lambda_{GB}$ is too large.
It is not clear that GB gravity has a sensible UV completion, 
but other theories where the KSS value is violated by a small amount do 
\cite{Kats:2007mq}.

A nice thing about the ratio $\eta \over s$ is that the number of degrees of freedom
($`\hl{N^2}$) cancels out.  Attempts to find other such observables, for example
related to charge transport, include \cite{Kovtun:2008kx}.

For lack of time, I've spoken here only about 
the extreme long-wavelength limit of the response functions.
The frequency and momentum dependence is also very revealing, 
see \eg\ \cite{Hartnoll:2007ih, Herzog:2007ij}.

\section{Concluding remarks}

\subsection{Remarks on other observables}

Besides correlation functions and thermodynamic potentials,
a number of other observables of interest can be computed easily
using the correspondence.
Some, like expectation values of Wilson loops, are 
relatively specific to gauge theories.

A very ubiquitous observable, which is notoriously hard to compute
otherwise, is the entanglement entropy.  
If we divide the hilbert space of the QFT into
\be\label{division}
\CH = \CH_A \oplus \CH_B \ee 
and declare ourselves ignorant of 
$\CH_B$, we have an ensemble described by the density matrix
\be \rho_A = \tr_{\CH_B} \rho \ee
where $\rho$ is the density matrix of the whole system,
which let's take to be the one associated with the ground state $\rho = \ket{\Omega}\bra{\Omega}$.
Then the entanglement entropy between $A$ and $B$ is 
the von Neumann entropy of $\rho_A$:
\be S = - \tr_A \rho_A \ln \rho_A .\ee

In the special case where the subdivision (\ref{division})
is done by cutting out a region of space $B$ bounded by some $d-2$-dimensional surface 
$\Sigma$, there exists a proposal for how to calculate the associated
entanglement entropy $S_\Sigma$ using the holographic dual 
\cite{Ryu:2006bv} (for a recent review, see \cite{Nishioka:2009un}).
The idea is extremely simple:
just find the surface $M$ ($d-1$-dimensional), ending on the boundary at $\Sigma$,
which extremizes its area.
The formula for $S_\Sigma$ is then
\be S_{\Sigma} = {\rm extremum}_{\partial_M = \Sigma} ~{{\rm area}(M)\over 4 G_N} ,\ee
very reminiscent of the Bekenstein formula for the entropy of a black hole.
This formula passes many checks.
For example, it gives the correct universal behavior
$S \sim {c\over 3} \ln L$ 
($L$ is the length of the region $B$) in a $1+1$-dimensional CFT of central  charge $c$.
Like the Casimir energy, this match between weak and strong coupling is precise because it 
is determined by the conformal anomaly.  
In higher dimensions, the holographic prescription 
gives a prediction for which terms in the expansion of $S(L)$ in powers of $L$ are universal.
There even exists a heuristic derivation 
\cite{Fursaev:2006ih}\footnote{
It should be noted that, at the moment at least,
there is some confusion about the case of two disconnected regions, see
\cite{Calabrese:2009ez}.}.


\subsection{Remarks on the role of supersymmetry}

Supersymmetry has played important roles in 
the historical development of the AdS/CFT correspondence:
\begin{itemize}
\item It constrains the form of the interactions, meaning that 
there are fewer possible candidates for the dual.
(For example, the maximal $AdS$ supergravity theory in five dimensions 
is unique.)

\item A supersymmetric theory has many observables which
are independent of the coupling.  These so-called BPS quantities
allow for many quantitative checks of a proposed dual pair.

\item Supersymmetry can stabilize a {\it line} of exact fixed points (for example 
in the $\CN=4$ SYM),
rendering the coupling constant a dimensionless parameter
which interpolates between the weakly-coupled field theory description
and the gravity regime.  
\end{itemize}

However, it has played no role in our discussion.
Some people believe that supersymmetry may be necessary for the 
construction of a consistent theory of quantum gravity.
But it seems more likely to me
that the formulation of specific examples of the duality 
without supersymmetry is a (perhaps hard) technical problem, not 
one of principle.


\subsection{Lessons for how to use AdS/CFT}

Critical exponents depend on `Landscape Issues'.  
By this I mean just that they depend on the values of the 
couplings in the bulk action (in particular, the masses of bulk fields), which 
are specified only by some UV completion, \ie\ by available string vacua.
For each possible bulk coupling, it is very much an open question 
which values arise in a consistent theory of quantum gravity.
This situation -- that the critical exponents are UV-sensitive quantities -- 
is a rather unfamiliar one!

At least in examples we know, thermodynamics is not 
very sensitive to strong coupling.  
In both the $\CN=4$ SYM, and from lattice QCD, 
we find for \eg\ the free energy a relation of the form 
\be F_{{\rm strong}} \sim {3\over 4} F_{{\rm weak}} .\ee

 Real-time dynamics and transport {\it are} very sensitive
to the strength of the interactions.  For example,
\be
 \({ \eta \over s} \)_{\rm weak} \sim { 1\over g^4 \ln g} ~~~~
\gg ~~~~  \({ \eta \over s} \)_{\rm strong} \sim {1\over 4\pi}  .\ee
Not only are these observables sensitive to strong coupling,
but they are very natural things to compute using the 
holographic technology.  In particular, 
although it is a classical description,
it automatically includes dissipation.
Ordinary techniques seem to require the existence
of a description of what's being transported in terms of quasiparticles, 
so that the Boltzmann equation
can be used.  Since we know that such a description needn't exist,
this is a very good opportunity for the machinery described above
to be useful.

I'd like to close with a final optimistic philosophical comment.
The following gedanken experiment was proposed by Weisskopf \cite{Weisskopf}:
Take a bunch of theoretical physicists 
and lock them away from birth so that they are never exposed to any
substance in the liquid phase.  
Will they predict the existence of the liquid phase from
first principles?
Weisskopf thinks not, because its existence depends on the fact that
the constituents interact strongly with each other.
The same statement applies to any state of matter
which depends for its existence on strong interactions,
such as confinement, and fractional quantum hall phases\footnote{Son points out that even some phenomena that only involve
weak coupling, such as the BCS mechanism, took a long time to figure out, even after the relevant experiments.}\footnote{I'd like to have more things to add to this list.  What am I forgetting?
Is our ignorance that complete?}.

I think it is a defensible claim \cite{HongLiu} that 
if we didn't know the IR physics of QCD (\eg\ if we didn't happen to be made out of 
color-neutral boundstates of quarks and gluons) before the discovery of AdS/CFT,
we would have predicted color confinement by finding its dual geometry 
\cite{Witten:1998zw, Klebanov:2000hb, Maldacena:2000yy}.
Our ability to imagine the possible behavior of a bunch of stuff 
has been limited by our dependence on our weak coupling tools, and 
on experimenters to actually assemble the stuff.
It is exciting that we now have another tool, 
which allows us to ask these questions in a way which involves
such simple geometrical pictures.  
Perhaps there are even states of matter that we can describe this way 
which have already been seen, but which have not yet been understood.

\vspace{0.2in}   \centerline{\bf{Other Acknowledgements}} \vspace{0.2in} 
Work supported in part by funds provided by the U.S. Department of Energy
(D.O.E.) under cooperative research agreement DE-FG0205ER41360,
and 
in part by the National Science Foundation under Grant No. NSF PHY05-51164.


\begin{thebibliography}{99}

\bibitem{Horowitz:2006ct}
  G.~T.~Horowitz and J.~Polchinski,
  ``Gauge/gravity duality,''
  arXiv:gr-qc/0602037.


  \bibitem{Son:2007vk}
 D.~T.~Son and A.~O.~Starinets,
``Viscosity, Black Holes, and Quantum Field Theory,''
 Ann.\ Rev.\ Nucl.\ Part.\ Sci.\  {\bf 57}, 95 (2007)
 [arXiv:0704.0240 [hep-th]].



\bibitem{Hartnoll:2009sz}
  S.~A.~Hartnoll,
  ``Lectures on holographic methods for condensed matter physics,''
  arXiv:0903.3246 [hep-th].

\bibitem{Sachdev:2008ba}
  S.~Sachdev and M.~Mueller,
  ``Quantum criticality and black holes,''
  arXiv:0810.3005 [cond-mat.str-el].

\bibitem{Maldacena:2003nj}
  J.~M.~Maldacena,
  ``TASI Lectures on AdS/CFT,''
  arXiv:hep-th/0309246.

\bibitem{Aharony:1999ti}
  O.~Aharony, S.~S.~Gubser, J.~M.~Maldacena, H.~Ooguri and Y.~Oz,
  ``Large N field theories, string theory and gravity,''
  Phys.\ Rept.\  {\bf 323}, 183 (2000)
  [arXiv:hep-th/9905111].

\bibitem{D'Hoker:2002aw}
  E.~D'Hoker and D.~Z.~Freedman,
  ``Supersymmetric gauge theories and the AdS/CFT correspondence,''
  arXiv:hep-th/0201253.

\bibitem{ocwref}
Course materials, including lecture notes and problem sets,
are available here: 
{\tt http://ocw.mit.edu/OcwWeb/Physics/8-821Fall-2008/CourseHome/index.htm}


\bibitem{Eisenbud}
D.~Eisenbud, {\it Commutative Algebra with a View Toward
Algebraic Geometry}, Springer 2004.


\bibitem{Jenkins:2006bz}
  S.~Weinberg and E.~Witten,
  ``Limits On Massless Particles,''
  Phys.\ Lett.\  B {\bf 96}, 59 (1980);
  A.~Jenkins,
  ``Topics in particle physics and cosmology beyond the standard model,''
  chapter 2,
  arXiv:hep-th/0607239.

\bibitem{Susskind:2005js}
  L.~Susskind and J.~Lindesay,
  {\it An introduction to black holes, information and the string theory
  revolution: The holographic universe,}
Hackensack, USA: World Scientific (2005) 183 p.


\bibitem{Bigatti:1999dp}
  D.~Bigatti and L.~Susskind,
  ``TASI lectures on the holographic principle,''
  arXiv:hep-th/0002044.

\bibitem{Maldacena:1997re}
  J.~M.~Maldacena,
  ``The large N limit of superconformal field theories and supergravity,''
  Adv.\ Theor.\ Math.\ Phys.\  {\bf 2}, 231 (1998)
  [Int.\ J.\ Theor.\ Phys.\  {\bf 38}, 1113 (1999)]
  [arXiv:hep-th/9711200].


\bibitem{Polchinski:1987dy}
  J.~Polchinski,
  ``Scale and conformal invariance in quantum field theory,''
  Nucl.\ Phys.\  B {\bf 303}, 226 (1988).


\bibitem{Swingle:2009bg}
B.~Swingle,
``Entanglement Renormalization and Holography,''
arXiv:0905.1317 [cond-mat.str-el].


  
\bibitem{Kachru:2008yh}
  S.~Kachru, X.~Liu and M.~Mulligan,
 ``Gravity Duals of Lifshitz-like Fixed Points,''
  Phys.\ Rev.\  D {\bf 78}, 106005 (2008)
  [arXiv:0808.1725 [hep-th]].
    
\bibitem{Son:2008ye}
  D.~T.~Son,
  ``Toward an AdS/cold atoms correspondence: a geometric realization of the
  Schroedinger symmetry,''
  Phys.\ Rev.\  D {\bf 78}, 046003 (2008)
  [arXiv:0804.3972 [hep-th]].
  
\bibitem{Balasubramanian:2008dm}
  K.~Balasubramanian and J.~McGreevy,
  ``Gravity duals for non-relativistic CFTs,''
  Phys.\ Rev.\ Lett.\  {\bf 101}, 061601 (2008)
  [arXiv:0804.4053 [hep-th]].

\bibitem{Klebanov:2000hb}
  I.~R.~Klebanov and M.~J.~Strassler,
  ``Supergravity and a confining gauge theory: Duality cascades and
  chiSB-resolution of naked singularities,''
  JHEP {\bf 0008}, 052 (2000)
  [arXiv:hep-th/0007191].




\bibitem{Susskind:1998dq}
  L.~Susskind and E.~Witten,
  ``The holographic bound in anti-de Sitter space,''
  arXiv:hep-th/9805114.


\bibitem{Silverstein:2003jp}
  E.~Silverstein,
  ``AdS and dS entropy from string junctions or the function of junction
  conjunctions,''
  arXiv:hep-th/0308175.



\bibitem{Belavin:1984vu}
  A.~A.~Belavin, A.~M.~Polyakov and A.~B.~Zamolodchikov,
  ``Infinite conformal symmetry in two-dimensional quantum field theory,''
  Nucl.\ Phys.\  B {\bf 241}, 333 (1984).



\bibitem{Heemskerk:2009pn}
  I.~Heemskerk, J.~Penedones, J.~Polchinski and J.~Sully,
  ``Holography from Conformal Field Theory,''
  arXiv:0907.0151 [hep-th].
  

\bibitem{Gubser:1998bc}
  S.~S.~Gubser, I.~R.~Klebanov and A.~M.~Polyakov,
  ``Gauge theory correlators from non-critical string theory,''
  Phys.\ Lett.\  B {\bf 428}, 105 (1998)
  [arXiv:hep-th/9802109].

\bibitem{Witten:1998qj}
  E.~Witten,
  ``Anti-de Sitter space and holography,''
  Adv.\ Theor.\ Math.\ Phys.\  {\bf 2}, 253 (1998)
  [arXiv:hep-th/9802150].



\bibitem{Silverstein:2004id}
  E.~Silverstein,
  ``TASI/PiTP/ISS lectures on moduli and microphysics,''
  arXiv:hep-th/0405068.
  
\bibitem{Douglas:2006es}
  M.~R.~Douglas and S.~Kachru,
 ``Flux compactification,''
  Rev.\ Mod.\ Phys.\  {\bf 79}, 733 (2007)
  [arXiv:hep-th/0610102].
  
\bibitem{Benvenuti:2004dy}
  S.~Benvenuti, S.~Franco, A.~Hanany, D.~Martelli and J.~Sparks,
  ``An infinite family of superconformal quiver gauge theories with
  Sasaki-Einstein duals,''
  JHEP {\bf 0506}, 064 (2005)
  [arXiv:hep-th/0411264].
  
  \bibitem{Coleman}
  S.~Coleman, `1/N'.
  
  \bibitem{ABJM}
  \eg\  J.~Bagger and N.~Lambert,
  ``Gauge Symmetry and Supersymmetry of Multiple M2-Branes,''
  Phys.\ Rev.\  D {\bf 77}, 065008 (2008)
  [arXiv:0711.0955 [hep-th]];
  O.~Aharony, O.~Bergman, D.~L.~Jafferis and J.~Maldacena,
  ``N=6 superconformal Chern-Simons-matter theories, M2-branes and their
  gravity duals,''
  JHEP {\bf 0810}, 091 (2008)
  [arXiv:0806.1218 [hep-th]],
  and citations thereof.

\bibitem{MatrixQM}
  P.~H.~Ginsparg and G.~W.~Moore,
  ``Lectures on 2-D gravity and 2-D string theory,''
  arXiv:hep-th/9304011.
    
\bibitem{Klebanov:2002ja}
  I.~R.~Klebanov and A.~M.~Polyakov,
  ``AdS dual of the critical O(N) vector model,''
  Phys.\ Lett.\  B {\bf 550}, 213 (2002)
  [arXiv:hep-th/0210114].
    
    
    \bibitem{probebranes}
  A.~Karch and E.~Katz,
``Adding Flavor to AdS/CFT,''
JHEP {\bf 0206} (2002) 043
[arXiv:hep-th/0205236].

\bibitem{probebraneapplications}
\eg\ 
A.~Karch, M.~Kulaxizi and A.~Parnachev,
``Notes on Properties of Holographic Matter,''
arXiv:0908.3493 [hep-th];
A.~Karch, D.~T.~Son and A.~O.~Starinets,
``Holographic Quantum Liquid,''
Phys.\ Rev.\ Lett.\  {\bf 102} (2009) 051602;
A.~Karch and A.~O'Bannon,
  ``Metallic AdS/CFT,''
  JHEP {\bf 0709}, 024 (2007)
  [arXiv:0705.3870 [hep-th]];  
  A.~Karch, D.~T.~Son and A.~O.~Starinets,
  ``Zero Sound from Holography,''
  arXiv:0806.3796 [hep-th];
  J.~L.~Davis, P.~Kraus and A.~Shah,
  ``Gravity Dual of a Quantum Hall Plateau Transition,''
  JHEP {\bf 0811}, 020 (2008)
  [arXiv:0809.1876 [hep-th]];
  E.~Keski-Vakkuri and P.~Kraus,
  ``Quantum Hall Effect in AdS/CFT,''
  JHEP {\bf 0809}, 130 (2008)
  [arXiv:0805.4643 [hep-th]];
    M.~Fujita, W.~Li, S.~Ryu and T.~Takayanagi,
  ``Fractional Quantum Hall Effect via Holography: Chern-Simons, Edge States,
  and Hierarchy,''
  JHEP {\bf 0906}, 066 (2009)
  [arXiv:0901.0924 [hep-th]];
  M.~Ammon, J.~Erdmenger, M.~Kaminski and P.~Kerner,
  ``Flavor Superconductivity from Gauge/Gravity Duality,''
  arXiv:0903.1864 [hep-th];
  J.~Erdmenger, N.~Evans, I.~Kirsch and E.~Threlfall,
``Mesons in Gauge/Gravity Duals - a Review,''
Eur.\ Phys.\ J.\  A {\bf 35} (2008) 81
[arXiv:0711.4467 [hep-th]].

    



  
\bibitem{Son:2002sd}
  D.~T.~Son and A.~O.~Starinets,
  ``Minkowski-space correlators in AdS/CFT correspondence: Recipe and
  applications,''
  JHEP {\bf 0209}, 042 (2002)
  [arXiv:hep-th/0205051].
  
  
    
\bibitem{Maldacena:2001kr}
  J.~M.~Maldacena,
  ``Eternal black holes in Anti-de-Sitter,''
  JHEP {\bf 0304}, 021 (2003)
  [arXiv:hep-th/0106112].
  
\bibitem{Herzog:2002pc}
  C.~P.~Herzog and D.~T.~Son,
  ``Schwinger-Keldysh propagators from AdS/CFT correspondence,''
  JHEP {\bf 0303}, 046 (2003)
  [arXiv:hep-th/0212072].
  
\bibitem{Skenderis:2008dh}
  K.~Skenderis and B.~C.~van Rees,
  ``Real-time gauge/gravity duality,''
  Phys.\ Rev.\ Lett.\  {\bf 101}, 081601 (2008)
  [arXiv:0805.0150 [hep-th]].
  
\bibitem{D'Hoker:1999pj}
E.~D'Hoker, D.~Z.~Freedman, S.~D.~Mathur, A.~Matusis and L.~Rastelli,
``Graviton Exchange and Complete 4-Point Functions in the AdS/CFT   Correspondence,''
Nucl.\ Phys.\  B {\bf 562} (1999) 353
[arXiv:hep-th/9903196].


  
\bibitem{Skenderis}
  \eg\ K.~Skenderis,
  ``Lecture notes on holographic renormalization,''
  Class.\ Quant.\ Grav.\  {\bf 19}, 5849 (2002)
  [arXiv:hep-th/0209067].
  
  


\bibitem{Henningson:1998gx}
M.~Henningson and K.~Skenderis,
``The Holographic Weyl Anomaly,''
JHEP {\bf 9807} (1998) 023
[arXiv:hep-th/9806087];
M.~Henningson and K.~Skenderis,
``Holography and the Weyl Anomaly,''
Fortsch.\ Phys.\  {\bf 48} (2000) 125
[arXiv:hep-th/9812032];
S.~de Haro, S.~N.~Solodukhin and K.~Skenderis,
``Holographic Reconstruction of Spacetime and Renormalization in the  AdS/CFT   Correspondence,''
Commun.\ Math.\ Phys.\  {\bf 217} (2001) 595
[arXiv:hep-th/0002230].


  
\bibitem{Balasubramanian:1999re}
  V.~Balasubramanian and P.~Kraus,
  ``A stress tensor for anti-de Sitter gravity,''
  Commun.\ Math.\ Phys.\  {\bf 208}, 413 (1999)
  [arXiv:hep-th/9902121].

\bibitem{Balasubramanian:1998de}
  V.~Balasubramanian, P.~Kraus, A.~E.~Lawrence and S.~P.~Trivedi,
  ``Holographic probes of anti-de Sitter space-times,''
  Phys.\ Rev.\  D {\bf 59}, 104021 (1999)
  [arXiv:hep-th/9808017].
  
  

\bibitem{Klebanov:1999tb}
  I.~R.~Klebanov and E.~Witten,
  ``AdS/CFT correspondence and symmetry breaking,''
  Nucl.\ Phys.\  B {\bf 556}, 89 (1999)
  [arXiv:hep-th/9905104].

\bibitem{Papadimitriou:2004ap}
I.~Papadimitriou and K.~Skenderis,
``AdS / CFT Correspondence and Geometry,''
arXiv:hep-th/0404176;
I.~Papadimitriou and K.~Skenderis,
``Correlation Functions in Holographic RG Flows,''
JHEP {\bf 0410} (2004) 075
[arXiv:hep-th/0407071].
  
\bibitem{Iqbal:2008by}
  N.~Iqbal and H.~Liu,
``Universality of the hydrodynamic limit in AdS/CFT and the membrane
paradigm,''
  Phys.\ Rev.\  D {\bf 79}, 025023 (2009)
  [arXiv:0809.3808 [hep-th]].

\bibitem{Iqbal:2009fd}
N.~Iqbal and H.~Liu,
``Real-Time Response in AdS/CFT with Application to Spinors,''
Fortsch.\ Phys.\  {\bf 57} (2009) 367
[arXiv:0903.2596 [hep-th]].

\bibitem{Bianchi:2001de}
M.~Bianchi, D.~Z.~Freedman and K.~Skenderis,
``How to Go with an RG Flow,''
JHEP {\bf 0108} (2001) 041
[arXiv:hep-th/0105276];
``Holographic Renormalization,''
Nucl.\ Phys.\  B {\bf 631} (2002) 159
[arXiv:hep-th/0112119];
K.~Skenderis and M.~Taylor,
``Kaluza-Klein Holography,''
JHEP {\bf 0605} (2006) 057
[arXiv:hep-th/0603016].



  
\bibitem{grawit}
  C.~R.~Graham and E.~Witten,
  ``Conformal anomaly of submanifold observables in AdS/CFT correspondence,''
  Nucl.\ Phys.\  B {\bf 546}, 52 (1999)
  [arXiv:hep-th/9901021].
  
\bibitem{shenker}
  L.~Fidkowski, V.~Hubeny, M.~Kleban and S.~Shenker,
  ``The black hole singularity in AdS/CFT,''
  JHEP {\bf 0402}, 014 (2004)
  [arXiv:hep-th/0306170].

\bibitem{Kraus:2002iv}
  P.~Kraus, H.~Ooguri and S.~Shenker,
  ``Inside the horizon with AdS/CFT,''
  Phys.\ Rev.\  D {\bf 67}, 124022 (2003)
  [arXiv:hep-th/0212277].
%

%

\bibitem{Hawking:1982dh}
  S.~W.~Hawking and D.~N.~Page,
  ``Thermodynamics Of Black Holes In Anti-De Sitter Space,''
  Commun.\ Math.\ Phys.\  {\bf 87}, 577 (1983).

\bibitem{Witten:1998zw}
  E.~Witten,
  ``Anti-de Sitter space, thermal phase transition, and confinement in gauge
  theories,''
  Adv.\ Theor.\ Math.\ Phys.\  {\bf 2}, 505 (1998)
  [arXiv:hep-th/9803131].
  
\bibitem{Maldacena:2000yy}
  J.~M.~Maldacena and C.~Nunez,
  ``Towards the large N limit of pure N = 1 super Yang Mills,''
  Phys.\ Rev.\ Lett.\  {\bf 86}, 588 (2001)
  [arXiv:hep-th/0008001].
  

  
\bibitem{Gubser:1996de}
A.~Strominger, unpublished;
  S.~S.~Gubser, I.~R.~Klebanov and A.~W.~Peet,
  ``Entropy and Temperature of Black 3-Branes,''
  Phys.\ Rev.\  D {\bf 54}, 3915 (1996)
  [arXiv:hep-th/9602135].
  
  \bibitem{latticeref}
\eg\
  G.~Boyd, J.~Engels, F.~Karsch, E.~Laermann, C.~Legeland, M.~Lutgemeier and B.~Petersson,
  ``Thermodynamics of SU(3) Lattice Gauge Theory,''
  Nucl.\ Phys.\  B {\bf 469}, 419 (1996)
  [arXiv:hep-lat/9602007].
  
  
  
   

\bibitem{Gubser:2008px}
  S.~S.~Gubser,
 ``Breaking an Abelian gauge symmetry near a black hole horizon,''
  Phys.\ Rev.\  D {\bf 78}, 065034 (2008)
  [arXiv:0801.2977 [hep-th]].
  
  
\bibitem{Hartnoll:2008vx}
  S.~A.~Hartnoll, C.~P.~Herzog and G.~T.~Horowitz,
  ``Building a Holographic Superconductor,''
  Phys.\ Rev.\ Lett.\  {\bf 101}, 031601 (2008)
  [arXiv:0803.3295 [hep-th]].
  
\bibitem{Hartnoll:2008kx}
  S.~A.~Hartnoll, C.~P.~Herzog and G.~T.~Horowitz,
 ``Holographic Superconductors,''
  JHEP {\bf 0812}, 015 (2008)
  [arXiv:0810.1563 [hep-th]].
  
\bibitem{Herzog:2009xv}
  C.~P.~Herzog,
  ``Lectures on Holographic Superfluidity and Superconductivity,''
  arXiv:0904.1975 [hep-th].
  
\bibitem{Lee:2008xf}
  S.~S.~Lee,
  arXiv:0809.3402 [hep-th].
  
\bibitem{LMV}
  H.~Liu, J.~McGreevy and D.~Vegh,
``Non-Fermi liquids from holography,''
  arXiv:0903.2477 [hep-th].


  
\bibitem{FLMV}
  T.~Faulkner, H.~Liu, J.~McGreevy and D.~Vegh,
``Emergent quantum criticality, Fermi surfaces, and $AdS_2$,''
arXiv:0907:2694 [hep-th].
  
  
\bibitem{Cubrovic:2009ye}
  M.~Cubrovic, J.~Zaanen and K.~Schalm,
  arXiv:0904.1993 [hep-th].

  
\bibitem{varma}
  C.~M.~Varma, P.~B.~Littlewood, S.~Schmitt-Rink, E.~Abrahams and A.~E.~Ruckenstein,
  ``Phenomenology of the normal state of Cu-O high-temperature
  superconductors,''
  Phys.\ Rev.\ Lett.\  {\bf 63}, 1996 (1989).

 
 
  
\bibitem{Horowitz:1999jd}
  G.~T.~Horowitz and V.~E.~Hubeny,
 ``Quasinormal modes of AdS black holes and the approach to thermal
 equilibrium,''
  Phys.\ Rev.\  D {\bf 62}, 024027 (2000)
  [arXiv:hep-th/9909056].

 
  
\bibitem{Buchel:2003tz}
  A.~Buchel and J.~T.~Liu,
  ``Universality of the shear viscosity in supergravity,''
  Phys.\ Rev.\ Lett.\  {\bf 93}, 090602 (2004)
  [arXiv:hep-th/0311175].
  
  
\bibitem{Kovtun:2003wp}
  P.~Kovtun, D.~T.~Son and A.~O.~Starinets,
  ``Holography and hydrodynamics: Diffusion on stretched horizons,''
  JHEP {\bf 0310}, 064 (2003)
  [arXiv:hep-th/0309213].

\bibitem{Kovtun:2004de}
  P.~Kovtun, D.~T.~Son and A.~O.~Starinets,
  ``Viscosity in strongly interacting quantum field theories from black hole
  physics,''
  Phys.\ Rev.\ Lett.\  {\bf 94}, 111601 (2005)
  [arXiv:hep-th/0405231].


\bibitem{Starinets:2008fb}
  A.~O.~Starinets,
  ``Quasinormal spectrum and the black hole membrane paradigm,''
  Phys.\ Lett.\  B {\bf 670}, 442 (2009)
  [arXiv:0806.3797 [hep-th]].

  
\bibitem{Schafer}  
    T.~Schafer and D.~Teaney,
  ``Nearly Perfect Fluidity: From Cold Atomic Gases to Hot Quark Gluon
  Plasmas,''
  arXiv:0904.3107 [hep-ph].

  
  \bibitem{ThomasVisc}
J.~Kinast, S.~L.~Hemmer, M.~E.~Gehm, A.~Turlapov, and
J.~E.~Thomas, Phys.\ Rev.\ Lett.\ {\bf 92}, 150402 (2004);
J.~Kinast, A.~Turlapov, J.~E.~Thomas, Phys.\ Rev.\ Lett.\
{\bf 94}, 170404 (2005) [cond-mat/0502507];
  B.~A.~Gelman, E.~V.~Shuryak and I.~Zahed,
  ``Cold Strongly Coupled Atoms Make a Near-perfect Liquid,''
  arXiv:nucl-th/0410067;
  T.~Schafer,
  ``The Shear Viscosity to Entropy Density Ratio of Trapped Fermions in the
  Unitarity Limit,''
  Phys.\ Rev.\  A {\bf 76}, 063618 (2007)
  [arXiv:cond-mat/0701251].

  
\bibitem{ShuryakEQ}
For reviews, see \eg\
  E.~Shuryak,
  ``Physics of Strongly coupled Quark-Gluon Plasma,''
  Prog.\ Part.\ Nucl.\ Phys.\  {\bf 62}, 48 (2009)
  [arXiv:0807.3033 [hep-ph]];
  W.~Busza,
  ``Lessons from PHOBOS,''
  arXiv:0907.4719 [nucl-ex].

  \bibitem{qgpreviews}
  S.~S.~Gubser and A.~Karch,
  ``From gauge-string duality to strong interactions: a Pedestrian's Guide,''
  arXiv:0901.0935 [hep-th];
    M.~P.~Heller, R.~A.~Janik and R.~Peschanski,
  ``Hydrodynamic Flow of the Quark-Gluon Plasma and Gauge/Gravity
  Correspondence,''
  Acta Phys.\ Polon.\  B {\bf 39}, 3183 (2008)
  [arXiv:0811.3113 [hep-th]];
  S.~S.~Gubser,
  ``Using string theory to study the quark-gluon plasma: progress and perils,''
  arXiv:0907.4808 [hep-th];
J.~Casalderrey-Solana, H.~Liu, D.~Mateos, K.~Rajagopal, U.~Wiedemann, to appear.
  



\bibitem{Brigante:2008gz}
  M.~Brigante, H.~Liu, R.~C.~Myers, S.~Shenker and S.~Yaida,
  Phys.\ Rev.\  D {\bf 77}, 126006 (2008)
  [arXiv:0712.0805 [hep-th]];
  ``The Viscosity Bound and Causality Violation,''
  Phys.\ Rev.\ Lett.\  {\bf 100}, 191601 (2008)
  [arXiv:0802.3318 [hep-th]].

\bibitem{Kats:2007mq}
  Y.~Kats and P.~Petrov,
  ``Effect of curvature squared corrections in AdS on the viscosity of the dual
  gauge theory,''
  JHEP {\bf 0901}, 044 (2009)
  [arXiv:0712.0743 [hep-th]].


  
\bibitem{Kovtun:2008kx}
  P.~Kovtun and A.~Ritz,
  ``Universal conductivity and central charges,''
  Phys.\ Rev.\  D {\bf 78}, 066009 (2008)
  [arXiv:0806.0110 [hep-th]].
  
    
\bibitem{Hartnoll:2007ih}
  S.~A.~Hartnoll, P.~K.~Kovtun, M.~Muller and S.~Sachdev,
  ``Theory of the Nernst effect near quantum phase transitions in condensed
  matter, and in dyonic black holes,''
  Phys.\ Rev.\  B {\bf 76}, 144502 (2007)
  [arXiv:0706.3215 [cond-mat.str-el]].
  
\bibitem{Herzog:2007ij}
  C.~P.~Herzog, P.~Kovtun, S.~Sachdev and D.~T.~Son,
  ``Quantum critical transport, duality, and M-theory,''
  Phys.\ Rev.\  D {\bf 75}, 085020 (2007)
  [arXiv:hep-th/0701036].


\bibitem{Ryu:2006bv}
  S.~Ryu and T.~Takayanagi,
  ``Holographic derivation of entanglement entropy from AdS/CFT,''
  Phys.\ Rev.\ Lett.\  {\bf 96}, 181602 (2006)
  [arXiv:hep-th/0603001].
  
  
\bibitem{Nishioka:2009un}
  T.~Nishioka, S.~Ryu and T.~Takayanagi,
  ``Holographic Entanglement Entropy: An Overview,''
  arXiv:0905.0932 [hep-th].
\bibitem{Fursaev:2006ih}
  D.~V.~Fursaev,
  ``Proof of the holographic formula for entanglement entropy,''
  JHEP {\bf 0609}, 018 (2006)
  [arXiv:hep-th/0606184].
  
\bibitem{Calabrese:2009ez}
  P.~Calabrese, J.~Cardy and E.~Tonni,
  ``Entanglement entropy of two disjoint intervals in conformal field theory,''
  arXiv:0905.2069 [hep-th].
  
  \bibitem{HongLiu}
  H.~Liu, private communication.
  
    
\bibitem{Weisskopf}
  H.~J.~Bernstein and V.~F.~Weisskopf,
  ``About liquids,'' MIT-CTP/1356.
  
\end{thebibliography}
\end{document}